\documentclass[a4paper,12pt]{article}
\usepackage{graphicx}
\usepackage{latexsym}
\usepackage{amsfonts}
\usepackage{amssymb}
\usepackage{amsmath}
\usepackage{psfrag}
\usepackage{rotating}
\usepackage[sc,small]{caption2}
\setcaptionwidth{14cm}

\numberwithin{equation}{section}

\begin{document}
\begin{flushright}
hep-th/0612224\\
AEI-2006-100\\
LMU-ASC 88/06\\
NA-DSF-36-2006 \\
NSF-KITP-2006-136
\end{flushright}
\vfill
\thispagestyle{empty}

\begin{center}
\textbf{\Large Glueballs vs.\ Gluinoballs:
Fluctuation Spectra in Non-AdS/Non-CFT}\\[48pt]%
Marcus Berg$^1$, Michael Haack$^2$ and Wolfgang M\"uck$^3$\\[24pt]%
\textit{$^1$ Max Planck Institute for Gravitational Physics }\\ %
\textit{ Albert Einstein Institute }\\  %
M\"uhlenberg 1, 14476 Potsdam, Germany \\[24pt] %
\textit{$^2$ Ludwig-Maximilians-Universit\"at}\\
Department f\"ur Physik, \\ 
Theresienstrasse 37, 80333 M\"unchen, Germany \\[24pt] %
\textit{$^3$ Dipartimento di Scienze Fisiche, Universit\`a degli Studi di
  Napoli ``Federico II''}\\%
\textit{and INFN, Sezione di Napoli}\\%
via Cintia, 80126 Napoli, Italy\\%
\end{center}
\vfill

\begin{abstract}
Building on earlier results on holographic bulk dynamics in confining gauge theories, 
we compute the spin-0 and spin-2 spectra of 
gauge theories dual to the non-singular Maldacena-Nunez and Klebanov-Strassler supergravity
backgrounds. We construct and apply a numerical recipe for
computing mass spectra from certain determinants. In the Klebanov-Strassler case, 
states containing the glueball and gluinoball
obey ``quadratic confinement'',
i.e.\ their mass-squareds depend on consecutive number as $m^2 \sim n^2$ 
for large $n$, with a universal proportionality constant.
The hardwall approximation appears to work poorly when compared to the unique spectra
we find in the full theory with a smooth cap-off in the infrared.
\end{abstract}
\vfill \vfill \vfill
\newpage


\newcommand{\ie}{i.e.,\ }
\newcommand{\eg}{e.g.,\ }


\newcommand{\rmd}{\,\mathrm{d}}

\newcommand{\Tr}{\operatorname{tr}}

\newcommand{\idmat}{\mathbb{I}}

\newcommand{\bzero}{\mathbf{0}}

\newcommand{\re}{\operatorname{Re}}
\newcommand{\im}{\operatorname{Im}}

\newcommand{\e}[1]{\operatorname{e}^{#1}}


\newcommand{\bp}{\bar{\phi}}
\newcommand{\bq}{\bar{q}}

\newcommand{\rir}{r_\text{IR}}

\newcommand{\rmid}{r_\text{mid}}

\newcommand{\vp}{\varphi}


\newcommand{\G}[2]{\mathcal{G}^{#1}_{\;\;#2}}

\newcommand{\mfa}{\mathfrak{a}}
\newcommand{\mfb}{\mathfrak{b}}
\newcommand{\mfe}{\mathfrak{e}}

\newcommand{\areg}{\mfa_{\mathrm{reg}}{}}
\newcommand{\asing}{\mfa_{\mathrm{sing}}{}}
\newcommand{\adom}{\mfa_{\mathrm{dom}}{}}
\newcommand{\asub}{\mfa_{\mathrm{sub}}{}}
\newcommand{\breg}{\mfb_{\mathrm{reg}}{}}
\newcommand{\bsing}{\mfb_{\mathrm{sing}}{}}
\newcommand{\bdom}{\mfb_{\mathrm{dom}}{}}
\newcommand{\bsub}{\mfb_{\mathrm{sub}}{}}

\newcommand{\phidom}{\phi_{\mathrm{dom}}{}}
\newcommand{\phisub}{\phi_{\mathrm{sub}}{}}

\newcommand{\adomzero}{\mfa_{\mathrm{dom}}^{(0)}{}}
\newcommand{\adomone}{\mfa_{\mathrm{dom}}^{(1)}{}}


\newcommand{\hrho}{\hat{\rho}}

\newcommand{\vev}[1]{\left\langle{#1}\right\rangle}

\newcommand{\bra}[1]{\langle{#1}|}
\newcommand{\ket}[1]{|{#1}\rangle}

\providecommand{\op}{\mathcal{O}}

\newcommand{\Order}[1]{\mathcal{O}\left(#1\right)}
\newcommand{\Of}{\Order{f}}
\newcommand{\Ofn}[1]{\Order{f^{#1}}}




\newcommand{\diag}{\operatorname{diag}}

\newcommand{\rmK}{\operatorname{K}}

\newcommand{\rmI}{\operatorname{I}}
\newcommand{\rmJ}{\operatorname{J}}

\newcommand{\rmP}{\operatorname{P}}



\newcommand{\tF}{\tilde{F}}

\newcommand{\N}{\mathcal{N}}

\newcommand{\D}{\mathcal{D}}

\newcommand{\cL}{\mathcal{L}}

\newcommand{\be}{\begin{equation}}
\newcommand{\ee}{\end{equation}}
\newcommand{\beqn}{\begin{eqnarray}}
\newcommand{\eeqn}{\end{eqnarray}}

\newcommand{\non}{\nonumber \\}
\newcommand{\hmm}[1]{{\bf [#1]}\marginpar[\hfill${\bf \Longrightarrow}$]%
                  {${\bf \Longleftarrow}$} }



\section{Introduction}
One of the first successes of 
gauge/string duality was the computation of mass spectra of strongly
coupled gauge theories
from dual geometries \cite{Witten:1998zw,Csaki:1998qr,deMelloKoch:1998qs}. 
These original papers were all concerned
with $\N=0$ black hole solutions, where it is hard
to check the validity of the correspondence. Other work uses the 
$\N=4$ superconformal theory (AdS bulk) plus a hard IR
cut-off \cite{Polchinski:2001tt} 
that serves to imitate interesting field theory IR effects, but there is no running coupling. 
Studying $\N=1$ super Yang-Mills theory coupled to matter 
seems a good compromise in many respects: these theories exhibit confinement, chiral
symmetry breaking, running couplings and a rich set of mass spectra. 

For this ``Non-AdS/Non-CFT correspondence'',\footnote{As far as we know,
  this expression was first used by Strassler at the Strings 2000
  conference. It also featured prominently in the titles of Aharony's
  2002 \cite{Aharony:2002up} and Zaffaroni's 2005 \cite{Zaffaroni:2005ty} lectures.}
where the bulk is not AdS and the
boundary field theory is not a CFT, much less is known about mass spectra, let
alone correlators. A few brave attempts exist in the literature 
\cite{Krasnitz:2000ir,Caceres:2000qe,Amador:2004pz,Caceres:2005yx}, but
as we explained in \cite{Berg:2005pd}, 
their results on mass spectra are at best inconclusive. 
In the present paper, we will compute the mass
spectrum of the $\N=1$ Klebanov-Strassler theory \cite{Klebanov:2000hb},
a non-conformal deformation of the $\N=1$ Klebanov-Witten theory \cite{Klebanov:1998hh} involving gluons and gluinos coupled to two sets of chiral superfields with a specific quartic
superpotential.\footnote{More precisely, we 
  consider only mass eigenstates that are dual to bulk solutions in the 
  Klebanov-Strassler background, i.e.\ we do not consider meson 
  mass spectra, which would arise from introducing flavor branes. 
  Moreover, there is always an issue with contamination by Kaluza-Klein states in these theories, which we will
  ignore in the following. For nice introductions to the 
  Klebanov-Strassler theory, see \cite{Herzog:2001xk,Strassler:2005}.}
We also compute the mass spectrum of the Maldacena-Nunez background 
\cite{Maldacena:2000yy}, as a warmup.
In this way, we will be able to address
physical  questions about how confinement works in
these models, such as whether the theory displays ``linear confinement''
\cite{Karch:2006pv}. 

Despite this promising state of affairs, 
many authors have emphasized that the aforementioned theories are
still quite far from real-world QCD. For instance, there are no
open strings corresponding to dynamical quarks, hence no real meson spectra. 
On the good side, in \cite{Sakai:2003wu,Kuperstein:2004hy}, probe D-branes were added to
the Klebanov-Strassler background, something which could be further developed using our methods.
Real QCD is also nonsupersymmetric, and we make heavy use of the existence of a ``superpotential'' $W$. 
However, as indicated by the quotation marks, this superpotential is ``fake" (in the sense of 
\cite{Freedman:2003ax}), but there are nonsupersymmetric examples where such structure exists 
irrespective of supersymmetry \emph{per se}, see e.g.\ \cite{Skenderis:2006jq}. Last but not least, real 
QCD is not at large 't Hooft coupling $\lambda$; this is very difficult to overcome with the present state
of the art, but 
the recent progress in \cite{Brower:2006ea} at least shows that $1/\sqrt{\lambda}$ corrections may 
not be unrealistic to obtain for these theories.

\begin{figure}[th]
\begin{center}
\includegraphics[width=0.9\textwidth]{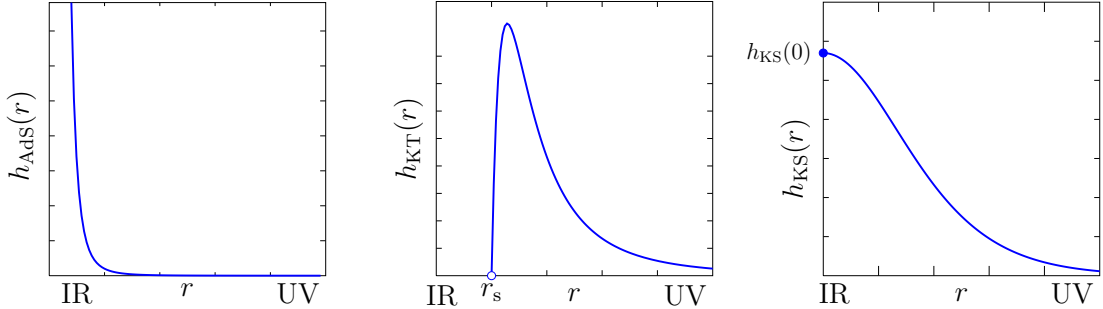}
\end{center}
\vspace{-5mm}
\caption{Warp factors $h(r)$. The string tension goes like $h^{-1/2}$.
From left to right the figures show the form of the warp factor in
AdS (tension goes to zero
in infrared), Klebanov-Tseytlin (tension diverges at $r_{\rm s}$) and
Klebanov-Strassler (tension goes to finite value).}
\label{hplots}
\end{figure}

Here, we take a step back from the effort to describe real QCD, and as mentioned above, focus on an example that is well-defined from a holographic point of view and see what can be understood about \emph{that} theory.
We introduce some new techniques, but conceptually the biggest difference from most of the literature is in things we do \emph{not} do. We do not employ the
``hard-wall'' approximation  (which amounts to taking AdS as in fig.~\ref{hplots} but with a
finite IR cutoff). 
One argument that has been put forward to motivate the use of the
hard-wall AdS model is that some physics should be insensitive to IR
details. However,
for computing mass spectra,
a boundary condition must be imposed in the IR, so this
question cannot really be insensitive to IR details (see e.g.\ \cite{Brower:2006ea}). 
We discuss to what extent it is in section \ref{sec:hardwall}.
Another common approximation, for example in the
context of inflation \cite{Baumann:2006th}, is the singular
Klebanov-Tseytlin background \cite{Klebanov:2000nc}, as in fig.\ \ref{hplots}.
We insist on using the full Klebanov-Strassler solution and imposing consistent boundary conditions, and find that  the commonly used approximations would likely not have led to correct results even for this theory, let alone for QCD. It seems reasonable to ask for explicit computational strategies to be developed concurrently with the search for real QCD, and 
we believe we have made progress on such strategies here.

It is important to recall that the Klebanov-Strassler theory
does not have a Wilsonian UV fixed point, but one can impose a
cutoff in the UV to define the theory. One can think of this in at
least three ways: 1) This is an intermediate approximation and the theory will be
embedded in a more complicated theory with a UV fixed point 
like in \cite{Hollowood:2004ek}; 2) we
will glue the noncompact dual geometry onto a compact space that naturally provides a UV
cutoff like in \cite{Giddings:2001yu}; or 
3) we will want to match to (at least lattice) data at that point, and beyond
that the theory is in any case not asymptotically free
(for recent ideas about this, see e.g.\ \cite{Evans:2006ea}).
As far as this paper is concerned, any of these points of view may be adopted.

Now to the new results in this paper.
We expand upon our earlier work \cite{Berg:2005pd} and present a numerical strategy, the ``determinant method'', to calculate the spectrum of regular and asymptotically subdominant bulk fluctuations.  
This puts us in the interesting position of being able to compute the
mass spectrum unambiguously, but not being able to say 
what the composition of each mass state is without further
information.\footnote{As will become clear, this is also true to some
  extent in standard AdS/CFT, but often ignored.}
 This information should ideally come from data,
i.e. one should compute mixings at a specified energy scale where one
has some control, and use the dual theory to evolve into the deep
nonperturbative regime. To illustrate this, let us take the simpler
example of a theory that does have a UV fixed point. The glueball operator
$\op_g$, with conformal dimension $\Delta=4$,
and the gluinoball operator $\op_{\tilde{g}}$, with $\Delta=3$, 
have a diagonal mass matrix in the
conformal theory. As soon as we let the theory flow away from conformality, there is a
nonzero correlator $\langle \op_g \op_{\tilde{g}} \rangle $. We can still
diagonalize at any given energy, and the mixing matrix will be energy
dependent. For theories that are not conformal even in the UV, there is no preferred basis
labelled by the $\Delta$ eigenvalues. One can still contemplate an
approximate labelling by $\Delta$ [in the KS theory, this corresponds
to expanding in the ratio $P=($number of fractional D3-branes$)/($number of
regular D3-branes$)$], but it is
not yet completely clear how to implement this in the dynamics;
as we pointed out in \cite{Berg:2005pd}, the limit $P\rightarrow 0$
does not commute with the UV limit one needs for the asymptotics.
We will give ideas about  this in sections \ref{method:polestruc} and \ref{sec:mixing}, but we
will not resolve it completely.

The spectra we find have some simple features. The states of the Klebanov-Strassler theory 
come in towers, each of which shows quadratic confinement for large excitation number
(i.e.\ $m^2 \sim n^2$, where 
$n$ denotes the excitation number within the tower). This confirms the claims of 
\cite{Karch:2006pv} also for the spin-0 states of the KS theory. However, 
for low excitation number (i.e.\ roughly for the first two excitations in each tower) 
the structure of the mass values is richer. 

For the Maldacena-Nunez background, the spectrum we find is rather different. It has an upper
bound, in agreement with the analytical spectra that we reported in \cite{Berg:2005pd}. 

Let us also mention that there are other approaches to holography in this
type of models, such as the Kaluza-Klein holography of \cite{Skenderis:2006uy}. Some issues may
appear in a different light in that framework, and that would be
interesting to know.

The organization of the paper is as follows. In chapter~\ref{method} we outline the general theory of bulk fluctuations as well as the correspondence between their spectrum and the spectrum of mass states in the dual gauge theory. The presentation is done such that it is applicable also in the case of 
non-asymptotically AdS bulk spaces. Based on this general material, a numerical strategy to calculate the specrum will be developed in chapter~\ref{mass_states}. 
As a warmup, our first application will be the Maldacena-Nunez 
theory \cite{Maldacena:2000yy} in chapter~\ref{MN}, before we come to the analysis 
of the mass spectrum in the Klebanov-Strassler cascading gauge theory in chapter~\ref{KS}. 
Some conclusions can be found in chapter~\ref{outlook}, and many of the more technical details 
have been deferred to the appendices.


\section{Holographic mass spectra}
\label{method}
We want to calculate mass spectra of confining gauge theories from the dynamics of fluctuations 
about their supergravity duals. In this section, we review and develop the main theoretical tools 
that are necessary for such a calculation. We will start, in Sec.~\ref{method:lindyn}, with a 
review of the treatment of the dynamics of supergravity 
fluctuations developed by us in \cite{Berg:2005pd}, 
generalizing a similar approach to the bulk
dynamics in AdS/CFT \cite{Bianchi:2003ug, Muck:2004qg, Muck:2004ih}. Then, 
in Sec.~\ref{method:findspec}, we identify the bulk duals of massive states in 
the gauge theory as eigenfunctions of a second-order differential operator 
satisfying suitable boundary conditions. Most of this material belongs to AdS/CFT 
folklore, but we will fill in some details not commonly found in the literature. 
Finally, in Sec.~\ref{method:polestruc}, we will attempt to justify the identification 
of the gauge theory mass spectrum with the spectrum of bulk eigenfunctions by 
studying the pole structure of holographic two-point functions. We hope that this 
goes some way towards a formulation of holographic renormalization
that holds beyond asymptotically AdS bulk spaces.

\subsection{Linearized bulk dynamics}
\label{method:lindyn}
Let us start by reviewing the methodology we employ for studying fluctuations about bulk 
backgrounds that are dual to confining gauge theories. We use consistent truncations of 
Type IIB supergravity that describe a subsector of the 10-dimensional dynamics in terms 
of a 5-dimensional effective theory. Details of this trunction can be found in our 
earlier paper \cite{Berg:2005pd}. Throughout this paper, we stick to a 5-dimensional 
effective bulk theory, but  generalization to arbitrary dimension is straightforward. 
The effective theory is a non-linear
sigma model of scalars coupled to 5-dimensional gravity, with an action of
``fake supergravity'' type,
\begin{equation}
\label{method:action5d}
  S = \int \rmd^5x \sqrt{g} \left[ - \frac14 R +\frac12
  G_{ab}(\phi) \partial_\mu \phi^a \partial^\mu \phi^b +V(\phi) \right]~.
\end{equation}
The ``fake supergravity'' property\footnote{Here ``fake'' does
not mean that the system is necessarily nonsupersymmetric --- the
systems
we considere here are supersymmetric --- just that the formalism is
applicable more generally. The relation 
between supergravity and fake supergravity was explored in \cite{Celi:2004st,Zagermann:2004ac}.} 
implies that the scalar potential $V(\phi)$ follows from a superpotential $W(\phi)$ by
\begin{equation}
\label{method:Vdef}
  V(\phi) = \frac12 G^{ab} W_a W_b -\frac43 W^2~.
\end{equation}
where we used the shorthand $W_a= \partial W/\partial \phi^a$.

Background solutions, also known as Poincar\'e-sliced domain walls or
holographic renormalization group flows, are of the form 
\begin{equation}
\label{method:bg1}
\begin{split}
  \rmd s^2 &= \rmd r^2 +\e{2A(r)} \eta_{ij} \rmd x^i \rmd x^j~,\\
  \phi^a &= \bp^a(r)~, 
\end{split}
\end{equation} 
where the background scalars, $\bp(r)$, and the warp function, $A(r)$,
are determined by the following first-order (BPS-like) differential equations,
\begin{equation}
\label{method:bg2}
\begin{split}
  \partial_r A(r) &= -\frac23 W(\bp)~,\\
  \partial_r \bp^a(r) &= W^a = G^{ab} W_b~.
\end{split}
\end{equation}

Fluctuations about the Poincar\'e-sliced domain wall backgrounds including
the scalar and metric fields are best described gauge-invariantly. We
will quote results from \cite{Berg:2005pd}. 
The physical fields are vectors in field space and scalars in spacetime, denoted by $\mfa^a$,
plus a traceless transverse tensor, $\mfe^i_j$, where $i,j=1,2,3,4$ are the indices along
4-dimensional Poincar\'e slices. These fields satisfy the following linearized field
equations,
\begin{equation}
\label{method:eqa}
  \left[ \left( \delta^a_b D_r +W^a_{\;\;|b} -\frac{W^aW_b}{W}
         -\frac83 W \delta^a_b \right) 
         \left( \delta^b_c D_r -W^b_{\;\;|c} +\frac{W^bW_c}{W} \right)
  +\delta^a_c \e{-2A} \Box \right] \mfa^c =0~,
\end{equation}
and
\begin{equation}
\label{method:eqe}
  \left( \partial_r^2 -\frac83 W \partial_r +\e{-2A} \Box
  \right) \mfe^i_j = 0~.
\end{equation}
In \eqref{method:eqa} and \eqref{method:eqe}, $\Box =\eta^{ij}
\partial_i \partial_j$, which will be replaced by $-k^2$ when we work
in 4-dimensional momentum space.  
Also, we use a sigma-model covariant notation, lowering and
raising indices of sigma-model tensors with $G_{ab}$ and its
inverse $G^{ab}$, and a
sigma-model covariant derivative $D_a$ that acts as
\begin{equation}
\label{method:cov.der}
  D_a \varphi_b \equiv \varphi_{a|b} \equiv \partial_b \varphi_a -\G{c}{ab} \varphi_c~,
\end{equation}
where $\G{c}{ab}$ is the Christoffel symbol derived from the metric $G_{ab}$. 
Furthermore, we define a ``background-covariant'' derivative $D_r$ ($r$ is the radial coordinate of \eqref{method:bg1}) acting as
\begin{equation}
\label{method:Dr}
  D_r \vp^a = \partial_r \vp^a + \G{a}{bc} W^b \vp^c~.
\end{equation}

So, our task is to solve
\eqref{method:eqa} and \eqref{method:eqe}
for the backgrounds of interest, imposing appropriate boundary
conditions that we now describe.


\subsection{Bulk duals of mass states}
\label{method:findspec}
In AdS/CFT, the usual definition of the bulk dual of a boundary gauge
theory mass state is a solution of the (linearized) bulk field
equations that is regular and integrable. 
In the following, we will state  these notions more precisely and present them in a form that is suitable also for application in non-AdS/non-CFT.

As prototype for the equations of motion \eqref{method:eqa} and
\eqref{method:eqe}, we consider a system of $n$ coupled second-order differential 
equations of the generic form
\begin{equation}
\label{method:eom.generic}
  \left[ D_r^2 -\frac83 W D_r +M(r) -\e{-2A(r)} k^2 \right] \mfa =0~,
\end{equation}
where $D_r$ is the background-covariant derivative \eqref{method:cov.der}, $M$ is a symmetric field-space tensor ($M_{ab}=M_{ba}$), and field indices have been omitted for the sake of brevity. 
We also assume  that the radial coordinate $r$ is defined in a domain
$\rir \leq r<\infty$, where $r\approx \rir$ corresponds to the deep
interior (IR) region, and large $r$ to the asymptotic (UV) region of
the bulk.\footnote{We allow $\rir=-\infty$, like for example in a
  pure AdS bulk.}


Let us start with the regularity condition. As we want the 10-dimensional configuration obtained via uplifting of an effective 5-dimensional solution to be singularity-free, we must impose that also the 5-d solutions be regular in the bulk.  For nonsingular backgrounds, \eqref{method:eom.generic} typically has a
regular singular point at $\rir$ and no others for finite $r$.\footnote{By ``typically'' we mean that
  this is the case in all regular holographic configurations known to us. It
  would be interesting to find more precise conditions for these
  statements.}
Regularity in the bulk, therefore, means regularity at $\rir$. To make
sure that our regularity condition is
invariant under changes of field space variables, we require that the norm of the fluctuation vector is finite at $r=\rir$, 
\begin{equation}
\label{method:regcond}
\text{\emph{Regularity condition:}} \qquad
  \left( \mfa^a G_{ab}\, \mfa^b \right)_{r=\rir} < \infty~. 
\end{equation}
We emphasize that one cannot simply demand regularity of the
components of $\mfa^a$ at $r=\rir$, because $r=\rir$ can coincide with a
coordinate singularity in the sigma-model metric $G_{ab}$, when evaluated on the background.
In the examples we consider, this indeed happens.

From the $2n$ independent solutions of \eqref{method:eom.generic}, the
condition \eqref{method:regcond}
typically\addtocounter{footnote}{-1}\footnotemark\ 
selects precisely $n$ independent regular solutions, which we will
denote\footnote{We hope there will be no confusion 
  between the index $i=1,\ldots ,n$ and the field space index $a$.} 
by $\areg_{,i}$, and $n$ singular solutions,
denoted by $\asing_{,i}$. 
A generic solution of \eqref{method:eom.generic} is a linear combination of these,
\begin{equation}
\label{method:gen0}
  \mfa(r) =  c_i\, \areg_{,i}(r) +\tilde{c}_i\, \asing_{,i}(r)  
\end{equation}
with constants $c_i$ and $\tilde{c}_i$. Thus, \eqref{method:regcond}
implies that 
\begin{equation}
  \tilde{c}_i = 0 \qquad \text{for all $i$.}
\end{equation}


Now, consider the integrability condition. Let us take a small detour and consider the  
bulk Green's function $G(r,r';k^2)$, which satisfies 
\begin{equation}
\label{method:eom.green} 
  \left[ D_r^2 -\frac83 W D_r +M(r) -\e{-2A(r)} k^2 \right] G(r,r';k^2) = 
 - \e{-4A(r)} \delta(r-r')~,  
\end{equation}
where the factor $\e{-4A}$ on the right hand side is the metric factor 
$1/\sqrt{g}$ from the covariant delta functional. The Green's
function can be written in terms of a basis of eigenfunctions, 
\begin{equation}
\label{method:green.ansatz} 
 G(r,r',k^2) = \sum\limits_\lambda 
  \frac{\mfa_\lambda(r) \mfa_\lambda(r')}{k^2 +m_\lambda^2}~,
\end{equation}
where the functions $\mfa_\lambda$ satisfy \eqref{method:eom.generic}
for $k^2= -m_\lambda^2$.(Again,
we omit the matrix indices, and the indices of the two $\mfa_\lambda$'s are not contracted.) 
Substituting \eqref{method:green.ansatz} into
\eqref{method:eom.green} yields the completeness relation
\begin{equation}
\label{method:complete} 
 \sum\limits_\lambda \mfa_\lambda(r) \mfa_\lambda(r') = 
  \e{-2A(r)} \delta(r-r')~,
\end{equation}
from which one can deduce the orthogonality relation
\begin{equation}
\label{method:ortho} 
 \int \rmd r\, \e{2A(r)} \mfa_\lambda(r)\cdot \mfa_\sigma(r) =
 \delta_{\lambda\sigma}~.
\end{equation}
With the dot product we denote the covariant contraction of indices. Eq.~\eqref{method:ortho} provides the
condition for the eigenstates $\mfa_\lambda$ to be integrable. Due to the factor $\e{2A}$ (for general dimension
$d$ it would be $\e{(d-2)A}$), the integral measure is not the covariant bulk integral measure that one might have expected.  

Now, the spectrum of bulk eigenfunctions $\mfa_\lambda$ is identified
with the mass spectrum of operators dual to the fields $\mfa$. 
(We will also say
that the eigenfunction $\mfa_\lambda$ is the bulk dual of a
state of mass $m_\lambda$.) 
Clearly, this spectrum depends on the boundary conditions one imposes. One boundary condition is given by the regularity condition \eqref{method:regcond}, the other follows from \eqref{method:ortho}, namely 
\begin{equation}
\label{method:intcond} 
\text{\emph{Integrability condition:}} \qquad
  \int \rmd r\, \e{2A(r)} \mfa(r)\cdot \mfa(r) <\infty~.
\end{equation}

For large $r$, \ie in the UV,
\eqref{method:eom.generic} has $2n$ independent asymptotic
 solutions. Let us denote the $n$ dominant
ones by $\adom_{,i}$, and the $n$ subdominant ones by $\asub_{,i}$ (by ``dominant'' we mean 
leading 
in $r$). 
Again, a generic solution of
\eqref{method:eom.generic} can be written as a linear combination of
these,
\begin{equation}
\label{method:genasymp}
  \mfa(r) = d_i \, \asub_{,i}(r) + \tilde{d}_i \, \adom_{,i}(r)\,  \; ,
\end{equation}
with constants $d_i$ and $\tilde{d}_i$.

It can be checked in the various cases we consider that the
asymptotically dominant behaviours, $\adom_{,i}$, are not integrable
in the sense of \eqref{method:intcond}, whereas the subdominant behaviours are integrable. 
Thus, \eqref{method:intcond} is equivalent to 
\begin{equation}
  \tilde{d}_i = 0 \qquad \text{for all $i$.}
\end{equation}

To summarize, the eigenfunctions $\mfa_\lambda$, interpreted as the bulk
duals of boundary field theory states with mass $m_\lambda$, are
such that 
\begin{equation}
\label{method:notildes}
  \tilde{c}_i= \tilde{d}_i = 0 \qquad \text{$i=1,2,\ldots n$,}
\end{equation}
from which follows that
\begin{equation}
\label{method:gendual}
  \mfa_\lambda(r) = c_{\lambda,i} \, \areg_{,i}(r) 
                  =  d_{\lambda,i} \, \asub_{,i}(r)~.
\end{equation}
%


\subsection{Pole structure of holographic 2-point functions}
\label{method:polestruc}

The masses of states (particles) in a quantum field theory 
manifest themselves as poles in the two-point functions
of operators, if there is a non-zero probability that these states are
created by the operators from the vacuum,\footnote{Typically, 
terms that are formally
  infinite and analytic in $k^2$ (contact terms, ``c.t.'') are
  needed to make the sum over the spectrum convergent. The finite
  parts of these counterterms are renormalization scheme
  dependent. For a
simple example, see Appendix~\ref{appads}.} 
\begin{equation}
\label{method:2ptQFT}
  \int \rmd^4 x\, \e{ikx} \vev{\op_1(x) \op_2(0)} = 
  \sum\limits_\lambda \frac{\bra{0} \op_1(0) \ket{\lambda} \bra{\lambda} \op_2(0) \ket{0}}{k^2 +m_\lambda^2}
   +\text{c.t.}\ .
\end{equation}
For comparison with this general formula, we will now try to obtain the pole structure of holographic 2-point functions from the linearized
dynamics of the bulk fields that we described in Sec.~\ref{method:lindyn}. But we
must start with a disclaimer. As we have not systematically dealt with
the renormalization and dictionary problems (see the discussion in
\cite{Berg:2005pd}), our results will depend on two strong
assumptions, motivated by how things work in AdS/CFT. A
better understanding of these assumptions as well as the meaning of
cases in which they are not satisfied would be very desirable. 

Let us start with the asymptotic expansion of the bulk fields
\eqref{method:genasymp}. For a given bulk field $\mfa$, the coefficients
$d_i$ and $\tilde{d}_i$ clearly depend on the choice of basis
functions $\asub_{,i}$ and $\adom_{,i}$, respectively. In order to remove
some of the ambiguities, we assume that the term containing $k^2$ in the
equation of motion \eqref{method:eom.generic} is asymptotically suppressed, so
that the leading terms of the asymptotic solutions are independent of
$k^2$, and the subleading terms can depend only on powers of
$k^2$ ({\it assumption 1}).\footnote{This assumption is quite strong, and as we shall see,
  it holds in the KS system, but not in the MN system.} 
Then, we can choose a basis of asymptotic solutions such that each has
a distinct leading behaviour. These solutions are classified according
to their asymptotic growth with the radial coordinate into dominant and subdominant solutions. 
With each dominant
asymptotic solution $\adom_{,i}$, we associate an operator $\op_i$ of
the dual field theory.\footnote{This does not make reference to \emph{components} of the field space vector
   $\mfa(r)$, which would require the operator $\op$ to carry a field index $a$ and make it transform under bulk field redefinitions.}
Two ambiguities still remain in the choice of basis functions, but we
now argue that these correspond to field theory ambiguities. 
First, the normalization of the leading terms in the asymptotic basis solutions
can be freely chosen. In field theory this corresponds to a choice of
normalization of the operators $\op_i$, which always drops out in
physical scattering amplitudes. Second, we have the freedom
to add to a given dominant solution multiples of asymptotic
solutions of equal or weaker strength, with coefficients polynomial in
$k^2$. (Solutions of equal strength can be added only with
coefficients independent of $k^2$.) 
This is also expected from the dual field theory, since operators of a given
dimension can mix with operators of equal or lower
dimensions (higher dimension operators being suppressed by our large
UV cutoff) which implies that generically, 
there is no unique way to define the renormalized operators. 
Therefore, the remaining ambiguities in the choice of an
asymptotic basis of solutions reflect the usual freedom in the
definition of field theory operators. 

For what follows, we do not need to make particular
choices for the subdominant basis solutions, $\asub_{,i}$. This
will be clearer once we obtain our final result of this section, i.e.\ (\ref{method:2ptfin}). 

Now for the second assumption.
In a field theory calculation of correlation functions of the operators
$\op_i$, one adds a source term to the Lagrangian,\footnote{The minus sign is a useful convention. Let $\vev{\op_i}_\text{exact}$ be the \emph{exact} 1-point function (\ie the 1-point function in the presence of finite sources $\tilde{d}_j$). Then, a connected $(n+1)$-point function involving $\op_i$ is given by the $n$-th derivative of $\vev{\op_i}_\text{exact}$ with respect to the sources $\tilde{d}_j$. With a plus sign, one would have an additional factor $(-1)^n$.}     
\begin{equation}
\label{method:duality}
 \Delta \cL = - \tilde{d}_i \op_i~.
\end{equation}
In holography, the sources are identified with the
coefficients $\tilde{d}_i$ of 
\eqref{method:genasymp}, whose dependence on $k^2$ was suppressed in that equation. 
The second strong assumption we make is the
generic form of the exact one-point function of the operators
$\op_i$. In AdS/CFT, the \emph{response} functions $d_i$ of
\eqref{method:genasymp} encode the exact one-point functions, up to scheme
dependent terms that are local in the sources 
\cite{Bianchi:2001kw, Martelli:2002sp, Papadimitriou:2004ap, Papadimitriou:2004rz}.\footnote{Imposing the
   regularity condition \eqref{method:regcond}, only one set of coefficients is independent, which is taken to be $\tilde{d}_i$.} 
In view of the ambiguities discussed above, we assume
\begin{equation}
\label{method:exact1pt}
  \vev{\op_i(k)}_{\text{exact}} = Y_{ij}(k^2) \, d_j(k^2) + \text{local terms}~,
\end{equation}
with a matrix $Y_{ij}(k^2)$, which we will be able to determine further
below ({\it assumption 2}). Part of the assumption is that the poles of the connected 2-point functions
arise in $d_j$, and that $Y_{ij}$ does not give rise to additional poles. 
In AdS/CFT, \eqref{method:exact1pt} 
follows from holographic renormalization, but unfortunately we have
no proof of it in this more general setting. 

We will now derive the general pole structure of holographic
2-point functions. To start, consider the general formula for a
solution $\mfa(r,k^2)$ of \eqref{method:eom.generic} in terms of the Green's
function and prescribed boundary values. 
Let $r_0$ be a (large) cut-off parameter determining the hypersurface
where the boundary values are formally prescribed. Remembering that
neither the Green's function nor its derivative vanish at the cut-off
boundary, we have\footnote{This formula follows from \eqref{method:eom.green} 
upon multiplication by $\e{4A(r)} \mfa(r)$ from the left, taking the integral over $r$, integrating by 
parts and using the field equation \eqref{method:eom.generic}. The IR boundary does not contribute, 
because $\e{4A}$ vanishes there. The reason for this is that $r=\rir$ should correspond 
to a single point of the bulk space, which is only 
guaranteed if $\e{4A}$ vanishes there, c.f.\ \eqref{method:bg1}.}
\begin{equation}
\label{method:gensol}
  \mfa(r,k^2) = \e{4A(r_0)} \left[ (D_{r_0} \mfa(r_0,k^2)) \cdot G(r_0,r;k^2) 
  - \mfa(r_0,k^2) \cdot D_{r_0} G(r_0,r;k^2)  \right]~,
\end{equation}
where $\mfa(r_0,k^2)$ and $D_{r_0} \mfa(r_0,k^2)$ are the prescribed values of
the field and its first derivative at the cut-off boundary, respectively.  
Since $r_0$ is an unphysical cut-off parameter, we must ensure that the
bulk field $\mfa(r)$ remains unchanged when $r_0$ is varied. This is
easily achieved, if, together with a change of the cut-off,
$r_0\to r_0 +\delta r_0$, the boundary values are changed by
\begin{equation}
\label{method:changebcs}
  \delta \mfa(r_0,k^2) = (D_{r_0} \mfa(r_0,k^2))\, \delta r_0~, \qquad
  \delta D_{r_0} \mfa(r_0,k^2) = (D^2_{r_0} \mfa(r_0,k^2))\, \delta
  r_0~,
\end{equation}
and the second derivative, $D^2_{r_0} \mfa(r_0,k^2)$, is determined by
the equation of motion \eqref{method:eom.generic}. To assure \eqref{method:changebcs}, we determine
the formal boundary values at the cut-off, $\mfa(r_0,k^2)$
and $D_{r_0} \mfa(r_0,k^2)$, from the generic asymptotic behaviour 
\eqref{method:genasymp}, with coefficients $d_i$ and $\tilde{d}_i$
fixed. After inserting \eqref{method:genasymp} and \eqref{method:green.ansatz} into
\eqref{method:gensol}, we obtain
\begin{equation}
\label{method:gensol2}
\begin{split} 
  \mfa(r,k^2) &= \e{4A(r_0)} \sum\limits_\lambda 
     \frac{\mfa_\lambda(r)}{k^2 +m_\lambda^2} \times \\
  &\quad \times \left\{ 
  d_l \left[ (D_{r_0}\asub_{,l}(r_0,k^2))\cdot \mfa_\lambda(r_0) 
     - \asub_{,l}(r_0,k^2)\cdot D_{r_0} \mfa_\lambda(r_0) \right]
     + \right. \\
  &\quad \quad + \tilde{d}_l \left. 
      \left[  (D_{r_0}\adom_{,l}(r_0,k^2))\cdot \mfa_\lambda(r_0)
      - \adom_{,l}(r_0,k^2)\cdot D_{r_0} \mfa_\lambda(r_0)
     \right] \right\}~.
\end{split}
\end{equation}
 
To continue, we observe that for very large $r_0$, the term on the
second line of \eqref{method:gensol2}, containing only subdominant
solutions, is much smaller than the term on the third line. Therefore,
we drop it. Moreover, as we are interested only in the pole structure,
we consider $k^2$ very close to $-m_\lambda^2$ and expand the numerator
keeping only the leading term, \ie we replace $k^2$ by $-m_\lambda^2$
in the numerator. Finally, we use the fact that the eigenfunctions are
purely subdominant,\footnote{At this point one may wonder where the
  dominant part of $\mfa$ comes from. It arises from the sum over
  the spectrum in \eqref{method:gensol2}, in particular from the UV
  contribution. For the simple case of AdS bulk, 
  we show this in Appendix \ref{appads}. However, it does not contribute to the poles.}
\begin{equation}
\label{method:eigen.subdom}
  \mfa_\lambda(r) = d_{\lambda,i}\, \asub_{,i}(r,-m_\lambda^2)~.
\end{equation}
This yields
\begin{equation}
\label{method:gensol3}
  \mfa(r,k^2) \underset{k^2 \approx -m_\lambda^2}{=} 
   \tilde{d}_l \, Z_{lj}(-m_\lambda^2) \, \frac{d_{\lambda,j}d_{\lambda,i}}{k^2 +m_\lambda^2}\;
   \asub_{,i}(r,-m_\lambda^2)~, 
\end{equation}
with
\begin{equation}
\label{method:Zdef}
  Z_{ij}(k^2) = \e{4A(r)} \left[ 
    (D_r \adom_{,i}(r,k^2))\cdot \asub_{,j}(r,k^2) -
    \adom_{,i}(r,k^2)\cdot D_r \asub_{,j}(r,k^2) \right]~.
\end{equation}
Notice that, by virtue of the equation of motion
\eqref{method:eom.generic}, $Z_{ij}$ is independent of $r$. 

Thus, after reading off the response function $d_i$ from
\eqref{method:gensol3}, we obtain the poles of 
the connected 2-point function, using \eqref{method:exact1pt} and
differentiating with respect to the source $\tilde{d}_j$,
\begin{equation}
\label{method:2pt}
  \vev{\op_i(k) \op_j (-k)} = \sum\limits_\lambda
  Y_{ii'}(-m_\lambda^2) Z_{jj'}(-m_\lambda^2) \, 
  \frac{d_{\lambda,i'}d_{\lambda,j'}}{k^2 +m_\lambda^2}
  + \text{c.t.}\ .
\end{equation}
By symmetry, we find that $Y_{ij}$ equals
$Z_{ij}$ up to normalization,
\begin{equation}
\label{method:Ydef}
    Y_{ij}(-m_\lambda^2) =  N_\lambda Z_{ij}(-m_\lambda^2)~.
\end{equation}
Therefore, defining also
\begin{align}
\notag
  Z_{\lambda,i} &= Z_{ii'} (-m_\lambda^2) \, d_{\lambda,i'} \\
\label{method:Zdef2}
  &= \e{4A(r)} \left[ (D_r \adom_{,i}(r,-m_\lambda^2))\cdot \mfa_\lambda(r) - 
   \adom_{,i}(r,-m_\lambda^2)\cdot D_r \mfa_\lambda(r)\right]~, 
\end{align}
our final result is 
\begin{equation}
\label{method:2ptfin}
  \vev{\op_i(k) \op_j (-k)} = \sum\limits_\lambda
  N_\lambda \frac{Z_{\lambda,i} Z_{\lambda,j}}{k^2 +m_\lambda^2}
  + \text{c.t.}\ .
\end{equation}
We note that the $Z_{\lambda,i}$ are independent of the choice of the
subdominant basis solutions, because the normalization of the
eigenfunctions $\mfa_\lambda$ is fixed by \eqref{method:ortho}. 


The determination of the normalization factors $N_\lambda$ is an open
problem, whose solution is intimately related with finding a proof of 
\eqref{method:exact1pt}. We conjecture that $N_\lambda =1$, as can be
easily checked for asymptotic AdS bulk spaces (see Appendix~\ref{appads}).
 


\section{Finding Mass States}
\label{mass_states}

In the previous section, we characterized the duals of mass states
as regular (in the IR) and asymptotically subdominant (in the UV) solutions of the linearized
bulk field equations. To calculate the spectrum, one solves a 
system of second-order differential equations
with boundary conditions specified at two endpoints. The standard
numerical strategy for this is ``shooting'', in
which one solves the corresponding initial value problem with two
boundary conditions (field value and derivative) specified at the
initial point, and then varies the initial condition trying to find
the desired boundary value at the final point. For coupled multifield
systems, this is a laborious process. Fortunately, since we are
interested only in the spectrum and not in the explicit dual bulk
solutions, we can adopt a much simpler strategy. It involves computing
a single
function, the determinant of a matrix, that depends on the
momentum $k^2$ and the vanishing of which signals the presence of a
mass state.

In Sec.~\ref{method:method1}, we shall present the general idea of our strategy in its simplest implementation involving the numerical solution of the field equations as an initial value problem starting close to $\rir$. 
This simple implementation presents some numerical challenges, which we describe, and which are overcome with the refined method described in Sec.~\ref{method:method2}.
Further details of numerical issues will be deferred to Appendix
\ref{app:numerics}.

\subsection{Mass Spectrum: Determinant Method}
\label{method:method1}

We begin by imposing regularity in the IR. This means that we consider 
the equations of motion for $r\approx \rir$, find independent solutions of these, and  
impose $\tilde{c}_i=0$ in the language of Sec.\ \ref{method:findspec}.
Starting with these initial conditions, one 
numerically evolves to larger $r$ to obtain the $n$ regular bulk
solutions $\areg_{,i}(r)$. 
For large $r$, the asymptotic behavior of each of these solution
can be expanded in terms of the UV-dominant solutions $\adom_{,j}$, so that
\begin{equation}
\label{method:regasymp}
  \areg_{,i}(r) \approx \gamma_{ji} \, \adom_{,j}(r) 
\qquad \text{(large $r$)}~,
\end{equation}
for some $k^2$-dependent but $r$-independent constants $\gamma_{ji}$. (The order of indices has been
chosen for later convenience. As a reminder, both $\areg_{,i}(r)$ 
and $\adom_{,j}(r)$ are $n$-component vectors in field space, but we
are suppressing the field-space index.)
Thus, a general regular solution 
(cf.\ \eqref{method:gen0})
will behave as 
\begin{equation}
\label{method:reggenasymp}
  \areg(r) \approx c_i \gamma_{ji} \, \adom_{,j}(r) 
\qquad \text{(large $r$)}~.
\end{equation}
In order for $\areg$ to qualify as the bulk dual of a boundary field
theory mass eigenstate, the remaining condition in
\eqref{method:notildes} is $\tilde{d}_i=0$ for the fluctuation to be
integrable. Thus,
\begin{equation}
\label{method:gammac}
  \tilde{d}_i=\gamma_{ij}c_j =0 \quad \text{for all $i$} 
  \quad \Longleftrightarrow \quad \text{mass state}~.
\end{equation}
Regarding this as a set of $n$ equations for the coefficients
$c_i$, it is necessary and sufficient for the existence of non-zero
solutions that the determinant of the matrix $\gamma_{ij}$ be
zero:
%
\begin{equation}
\label{method:asympcond}
  \det \gamma_{ij}(k^2) = 0
  \quad \Longleftrightarrow \quad \text{mass state with $m^2=-k^2$}~,
\end{equation}
where we have restored the dependence of
the matrix $\gamma_{ij}$ on the 4-momentum $k^2$.

To compute $\gamma_{ij}$, let us make the field-space component
index explicit in eq.\ \eqref{method:regasymp}:
\begin{equation}
\label{method:regasymp2}
  (\areg(z))_{~i}^a \approx \gamma_{ji} (\adom(z))_{~j}^a 
\end{equation}
where $\areg(z)$ and $\adom(z)$ are viewed as $n\times n$ matrices,
with column index $i$ and row index $a$. 
Inverting this equation, the coefficient matrix 
$\gamma_{ij}$ can be found by calculating\footnote{This equation is
  intended in the sense of matrix multiplication, so there 
is no $G_{ab}$ involved.}
\begin{equation}
\label{method:gamma}
  \gamma_{ij} \approx \left[(\adom)^{-1} (\areg) \right]_{ij} 
\end{equation}
for some large value of $r$, where the subdominant parts of 
$\areg$ are sufficiently suppressed. 
We would like to emphasize that formula \eqref{method:gamma} 
renders the coefficients
$\gamma_{ij}$ independent of the coordinates in field space 
--- the component index $a$ is
contracted. Hence, also the mass condition \eqref{method:asympcond}
is independent of the choice of sigma-model variables.

However, as announced above, 
the IR $\rightarrow$ UV determinant method does not work straightforwardly
for coupled multifield systems in practice. The method is only
sufficiently insensitive to numerical error
if all dominant solutions $\adom_{,i}$ are comparable in
magnitude, such that the integration error from the stronger ones does
not swamp the significant digits of the weaker ones. 
For 1-component systems, this is obviously never a problem, and it
turns out also not to be a problem for the
2-component system in the MN background (Section~\ref{MN:numeric}). 
In general, and in particular for the 
7-component spin-0 system in the
KS background, 
it is better to tread more carefully and only evolve to a prescribed
midpoint value, as we will describe in the next subsection.

One could have imagined that this numerical difficulty would be alleviated
by evolving the other way (UV $\rightarrow$ IR). This is actually
partially true, but the method in the next subsection is still more
useful (see Appendix \ref{app:numerics} for a few more details).
Indeed, imposing the relevant
initial conditions at a large value of $r$, one can calculate 
numerically the $n$ UV-subdominant solutions $\asub_{,i}$.
Decomposing their behavior at $r\approx \rir$ using \eqref{method:gen0}
into 
\begin{equation}
\label{method:sub0}
  \asub_{,i}(r) = \gamma_{ji}\, \asing_{,j}(r)  
  + {\gamma'}_{ji} \, \areg_{,j}(r)~,
\end{equation}
the same argument as before affirms that an $\asub$ dual to a mass
eigenstate exists, if $\det \gamma_{ij}=0$. The coefficients
$\gamma_{ij}$ can be determined by matching the components of 
$\asub$ to the generic singular behaviour. 

The build-up and growth of integration error on the way from the asymptotic
region to $\rir$ is an issue here and must be resolved. This can be
done by making sure the UV cutoff is not {\it too} large. Then, the
integration error, which generally grows like the asymptotically 
weakest solution, remains small.

\subsection{Mass Spectrum: Midpoint Determinant Method}
\label{method:method2}

The midpoint method is numerically the most stable and combines the two previous approaches. 
One calculates the asymptotic solutions analytically both in the UV and in the IR.
Then, the $n$ regular solutions $\areg_{,i}$ are evolved
from $r\approx \rir$ up to a mid-point $\rmid$, and the $n$ subdominant
solutions $\asub_{,i}$ are evolved from large $r$ down to $\rmid$. 
Then, one tries to find linear combinations of the regular
solutions and of the subdominant solutions, respectively, such that
their value and first derivative at $\rmid$ match:
%
\begin{equation}
\label{method:mid-point}
 \begin{pmatrix} \areg & \asub \\
      \partial_r \areg & \partial_r \asub \end{pmatrix}_{\!\!r=\rmid} 
  \begin{pmatrix} c \\ -d \end{pmatrix} = 0~.
\end{equation}
As before, the necessary and sufficient condition for the existence of
a mass eigenstate is given by the determinant of a matrix (this time
$2n \times 2n$) being zero.
Put differently, rather than demanding that a given individual
field match smoothly 
across the midpoint, we allow for the situation that only some linear combination of the basis
fields matches smoothly, and this linear combination is
determined by the coefficients $c$ and $d$.

If the solutions we are matching were approximate analytical
solutions and the midpoint were the classical turning point, 
this midpoint matching would be precisely the WKB approximation. Of course,
our solutions will not be approximate analytical, but numerical and exact
(up to integration error), so it is no approximation in that sense.

\section{Maldacena-Nunez system}
\label{MN}

\subsection{General Relations}
\label{MN:bg}

The effective 5-d model describing the bulk dynamics of the MN system
contains three scalar fields $(g,a,p)$ and is characterized by the
sigma-model metric \cite{Papadopoulos:2000gj,Berg:2005pd}
\begin{equation}
\label{MN:G}
  G_{ab} \partial_\mu \phi^a \partial^\mu \phi^b = 
  \partial_\mu g \partial^\mu g 
  + \e{-2g} \partial_\mu a \partial^\mu a 
  + 24 \partial_\mu p \partial^\mu p~,
\end{equation}
and the superpotential\footnote{We have adjusted the overall factor of
  the superpotential of \cite{Papadopoulos:2000gj} to our conventions.}
\begin{equation}
\label{MN:W}
  W = -\frac12 \e{4p} \left[ (a^2-1)^2 \e{-4g} +2(a^2+1) \e{-2g} +1
    \right]^{1/2}~.
\end{equation}

Let us summarize what 
Poincar\'e-sliced domain wall backgrounds this system admits. 
In what follows, we shall denote the background
values of the scalar fields by $g$, $a$ and $p$, and for their
fluctuations we will use the gauge-invariant sigma-model vector
$\mfa=(\delta g, \delta a, \delta p)^T$.\footnote{The superscript $T$ 
denotes the transpose, because $\mfa$ is a column vector.} The
background equations \eqref{method:bg2} are most easily solved in
terms of a new radial coordinate, $\rho$, defined by
\begin{equation}
\label{MN:rhodef}
  \partial_\rho = 2 \e{-4p} \partial_r~.
\end{equation}
We shall need only the explicit solutions for $a$ and $g$, which are
given by
\begin{equation}
\label{MN:ag.sol}
  a = \frac{2\rho}{\sinh(2\rho+c)}~,\quad 
  \e{2g} =4\rho\coth(2\rho+c) -(a^2+1)~.
\end{equation}
The integration constant $c$, which appeared for the first time in
\cite{Gubser:2001eg}, can take values $0\leq c \leq \infty$. The regular
MN solution corresponds to $c=0$, while $c>0$ leads to singular bulk
geometries. The solution \eqref{MN:ag.sol} is defined for $\rho\geq
\hrho$, with $\hrho$ defined by $\e{2g(\hrho)}=0$, from
which one obtains the relation
\begin{equation}
\label{MN:hrho}
  \coth(2\hrho +c) =\frac1{2\hrho}-1~.
\end{equation}
Hence, $\hrho$ takes values between $0$ and $1/4$, with $\hrho=0$ for
$c=0$ and $\hrho=1/4$ for $c=\infty$. For $c>0$, $\rho=\hrho$ is the
location of the singularity.

For the scalar $p$ we will only need its relation to the warp function
$A$ (see \eqref{method:bg1}), which is 
\begin{equation}
\label{MN:pA.rel}
  \e{-2A} \e{-8p} =C^2~,
\end{equation}
where $C^2$ is an integration constant setting the 4-d scale. For
later convenience, we set $C^2=1/4$. 

Let us turn to the fluctuation equations \eqref{method:eqa} and
\eqref{method:eqe} in terms of the
radial variable $\rho$. After using the relations for the background,
one straightforwardly finds that \eqref{method:eqa} becomes
\begin{equation}
\label{MN:eqa}
  \left[ (\partial_\rho - M)(\partial_\rho -  N) - k^2 \right] \mfa
  =0~,
\end{equation}
where we dropped field indices,
and the matrices $M$ and $N$ are given by
\begin{equation}
\label{MN:mats}
\begin{split}
  N^a_{~b} &=2\e{-4p} \left(\partial_b W^a -\frac{W^a W_b}{W} \right)~,\\
  M^a_{~b} &= -N^a_{~b} -4 \e{-4p} \left(\G{a}{bc}W^c -W \delta^a_b\right)~.
\end{split}
\end{equation}
Note that $M$ and $N$ are independent of $p$. 
As was observed in \cite{Berg:2005pd},
the fluctuations of $p$, which are described by the
component $\mfa^3$, decouple from the other two. This can be seen from
\eqref{MN:mats} as follows. From \eqref{MN:G} and \eqref{MN:W} we find
the identities 
\begin{equation}
\label{MN:W3.ident}
  W_3 = 4W~,\qquad W^3 = \frac16 W~,\qquad \partial_p W^a = 4 W^a~,
\end{equation}
from which follow $N^a_{~3}=N^3_{~b}=0$. Moreover, one easily checks that
$\G{3}{bc}=\G{a}{3c}=0$, so that $M^a_{~3}=4\e{-4p}W\delta^a_3$, and
$M^3_{~b}=4 \e{-4p}W\delta^3_b$. Thus, $\mfa^3$ satisfies 
\begin{equation}
\label{MN:a3.eq}
  \left[ (\partial_\rho -4\e{-4p}W) \partial_\rho - k^2 \right]
  \mfa^3 =0~.
\end{equation}
This is just the equation for a massless scalar field in the domain
wall background. 

Similarly, \eqref{method:eqe} gives rise to 
\begin{equation}
\label{MN:eqe}
  \left[ (\partial_\rho -4\e{-4p}W) \partial_\rho - k^2 \right]
  \mfe^i_j =0~.
\end{equation}
which is identical to \eqref{MN:a3.eq}.

\subsection{Analytic solutions in singular background}
\label{MN:analytic}

We will briefly introduce the analytical
solutions for the fluctuations in the singular background case
$c=\infty$ derived in \cite{Berg:2005pd}. These analytical solutions
give useful indications for  glueball and gluinoball
masses that can be compared with the numerical analysis in
Sec.~\ref{MN:numeric}. Further details can be found in our earlier paper
\cite{Berg:2005pd}.

Let us start with \eqref{MN:eqa}. In the case $c=\infty$, the matrices
$M$ and $N$ are very simple and diagonal,
\begin{equation}
\label{MN:MN.asymp}
\begin{split}
  M &= \frac1{4\rho-1} \diag \left( -8\rho +4 -\frac1{\rho},
  -2+\frac1{\rho}, -8\rho \right)~,\\
  N &= \diag \left( -\frac1{\rho}, \frac1{\rho}-2,0 \right)~.
\end{split}
\end{equation}
With these expressions, \eqref{MN:eqa} can be solved
analytically. We shall illustrate the analysis considering the
independent solutions for the second component,
\begin{equation}
\label{MN:a2}
  \mfa^2 \sim \e{-(\alpha+1/2)z} 
  \begin{cases}
    (\alpha z)^{3/2}\, 
    \Phi\left( \frac54-\frac32\alpha,\frac52;z\right)~,\\
    \Phi\left(-\frac14-\frac32\alpha,-\frac12;z\right)~,
  \end{cases}
\end{equation}
where $\Phi$ is the confluent hypergeometric function of the first kind (or Kummer
function $\mathrm{M}$), and $\alpha$ and $z$ are defined
by\footnote{The combination $\rho-1/4$ is simply $\rho
  -\hrho$; see the discussion around \eqref{MN:hrho}.}  
\begin{equation}
\label{MN:alphazdef}
  \alpha= \frac12 (1+k^2)^{-1} ~,\qquad 
  z = \frac1{\alpha} \left(\rho-\frac14\right)~.
\end{equation}

The two solutions in \eqref{MN:a2} have different
behaviour at the singularity, \ie for $\rho\approx \hrho=1/4$. Notice that both solutions satisfy the regularity condition
\eqref{method:regcond}, so we do not have the standard means of
excluding one of them. In other words, in the background with
$c=\infty$, the regularity condition \eqref{method:regcond} does not
uniquely specify the IR boundary value that should be
imposed on the eigenfunctions.\footnote{One can show that the same
  behaviour is found for all $c>0$ by expanding the matrices $M$ and
  $N$ for $\rho\approx \hrho$. The regular case $c=0$ is
  qualitatively different from these, because the limits $c\to 0$ and
  $\rho\to \hrho$ do not always commute. To see this, consider, \eg $\rmd
  a(\rho)/\rmd \rho$ at $\rho=\hrho$.}  
Let us tentatively discard the solution with the leading small-$z$ behaviour of
\eqref{MN:a2}, which will be justified by the numerical results in the next section. 
This leaves us with the first solution. The generic
asymptotically dominant UV behaviour is absent in this solution, if
$\alpha$ takes one of the values \cite{Berg:2005pd} 
\begin{equation}
\label{MN:spec1} 
  \alpha = \frac16 (4n+1)~,\qquad n=1,2,3,\ldots~.
\end{equation}
As the second component is dual to the gluino bilinear, these states
may be interpreted as gluinoballs. 

Applying the same argument to the solutions of the first component
\cite{Berg:2005pd}, we find states for 
\begin{equation}
\label{MN:spec2} 
  \alpha = \frac16 (4n+3)~, \qquad n=1,2,3,\ldots~.
\end{equation}

The spectra \eqref{MN:spec1} and \eqref{MN:spec2} can be combined into
\begin{equation}
\label{MN:spec3} 
  \alpha = \frac16 (2n+3)~, \qquad n=1,2,3,\ldots~,
\end{equation}
with odd and even $n$ corresponding to \eqref{MN:spec1} and
\eqref{MN:spec2}, respectively. Using \eqref{MN:alphazdef}, we can
rewrite \eqref{MN:spec3} as
\begin{equation}
\label{MN:spec3m} 
  m^2 = \frac{4n(n+3)}{(2n+3)^2}= 1 -\frac{9}{(2n+3)^2}~, \qquad n=1,2,3,\ldots~.
\end{equation}
We would like to remark that, in \cite{Berg:2005pd}, we had obtained a
spectrum identical to \eqref{MN:spec3m}, with the addition of a
massless gluinoball for $n=0$. The above heuristic argument,
that the solution with leading small-$z$ behavior should be discarded, disposes of the massless
gluinoball. This choice will be justified \emph{a posteriori} by the
numerical results.
 
As we move from the background with $c=\infty$ to the regular
background with $c=0$, we expect these values to change. Moreover,
in the regular background, no heuristic argument is needed, as the
regularity condition will exclude half the solutions.

From the solutions of the third component, no discrete mass values are
found \cite{Berg:2005pd}.

\subsection{Numerical solutions in the regular background}
\label{MN:numeric}

Let us now consider  \eqref{MN:eqa} in the regular background, \ie for
$c=0$. The matrices $M$ and $N$ are considerably more complicated than
in the previous case, and we list their entries in
Appendix~\ref{app:MNmats}. Thus, the full equations of motion will be
solved numerically.

For our purposes, the simple approach of Sec.~\ref{method:method1} is
sufficient for the MN solution. (It will not be sufficient 
for the KS solution.) Hence, let us first consider $\rho$ in the vicinity of
$\hrho=0$, and find the regular solutions of \eqref{MN:eqa}.
Expanding the matrices $M$ and $N$ about $\rho=0$ yields
\begin{align}
\label{MN:M.expand}
  M &= \begin{pmatrix}  
          -\frac2{\rho}+\frac{4\rho}3 & \frac1{\rho}-\frac{34\rho}{27}
          & 0 \\
          \frac{4\rho}3 & \frac1{\rho} -\frac{8\rho}3 & 0 \\
          0 & 0 & -\frac2{\rho} -\frac{8\rho}9 
       \end{pmatrix} +\Order{\rho^3}~,\\
\label{MN:N.expand}
  N &= \begin{pmatrix}  
          -\frac{20\rho}9 & -\frac1{3\rho}+\frac{14\rho}{15}
          & 0 \\
          -4 \rho & -\frac1{\rho} +\frac{8\rho}9 & 0 \\
          0 & 0 & 0 
       \end{pmatrix} +\Order{\rho^3}~.
\end{align}
Recall from the discussion at the end of Sec.~\ref{MN:bg} that the
zeros in \eqref{MN:M.expand} and \eqref{MN:N.expand} are exact. 

Using \eqref{MN:M.expand} and \eqref{MN:N.expand}, it is
straightforward to solve \eqref{MN:eqa} in terms of 
series expansions about $\rho=0$. The independent regular behaviours are 
\begin{align}
\label{MN:a.reg1}
  \mfa_1 &= \left[ 1+\frac16\left( k^2 -\frac83\right) \rho^2 \right] 
           \begin{pmatrix} 1\\0\\0 \end{pmatrix} +\Order{\rho^4}~,\\ 
\label{MN:a.reg2}
  \mfa_2 &= \rho^2
           \begin{pmatrix} 1\\3\\0 \end{pmatrix} +\Order{\rho^4}~,\\ 
\label{MN:a.reg3}
  \mfa_3 &= \left( 1+\frac16 k^2 \rho^2 \right)
           \begin{pmatrix} 0\\0\\1 \end{pmatrix} +\Order{\rho^4}~.
\end{align}
The other three solutions go as $1/\rho$ in the leading term and will
be discarded. 

The asymptotic region ($\rho\gg 1$) is well described by the singular
background considered in Sec.~\ref{MN:analytic}. Hence, a convenient
basis of generic dominant asymptotic behaviours can be found from the 
analytical solutions of \cite{Berg:2005pd}. 
Combining them to a matrix, where each column is an independent
solution, we get
\begin{equation}
\label{MN:asymp.matrix}
  \adom = \e{-\left(1-{1 \over 2\alpha}\right)\rho} 
  \begin{pmatrix}
    \rho^{-1/4-3\alpha/2} & 0 & 0 \\
    0 & \rho^{1/4-3\alpha/2} & 0 \\
    0 & 0 & \rho^{-1/4+\alpha/2} 
  \end{pmatrix}~.
\end{equation}

%
%
\begin{table}[th]
\caption{Comparison between numerical results ($m^2$) and analytical estimates
  for the mass spectrum in the MN background.\label{MN:masstable}}
\begin{center}
\begin{tabular}{|r|c|c||r|c|c||r|c|c|}
\hline
$n$ & $m^2$ & \eqref{MN:spec3m} & 
$n$ & $m^2$ & \eqref{MN:spec3m} & 
$n$ & $m^2$ & \eqref{MN:spec3m} \\
\hline
1  & 0.4068 & 0.6400  &
9  & 0.9782 & 0.9796  &
17 & 0.9932 & 0.9934  \\
2  & 0.8078 & 0.8163  &
10 & 0.9828 & 0.9830  &
18 & 0.9940 & 0.9941  \\
3  & 0.8675 & 0.8889  &
11 & 0.9848 & 0.9856  & 
19 & 0.9945 & 0.9946  \\
4  & 0.9235 & 0.9256  &
12 & 0.9875 & 0.9877  & 
20 & 0.9951 & 0.9951  \\
5  & 0.9406 & 0.9467  & 
13 & 0.9888 & 0.9893  &
21 & 0.9954 & 0.9956  \\
6  & 0.9592 & 0.9600  &
14 & 0.9905 & 0.9906  &
22 & 0.9959 & 0.9959  \\
7  & 0.9662 & 0.9689  & 
15 & 0.9914 & 0.9917  & 
23 & 0.9961 & 0.9962  \\
8  & 0.9747 & 0.9751  &
16 & 0.9926 & 0.9927  &
24 & 0.9965 & 0.9965  \\
\hline
\end{tabular}
\end{center}
\end{table}   

Putting into practice the method of Sec.~\ref{method:method1}, we have
searched for zeros of $\det \gamma$, where the matrix $\gamma$ is
defined by \eqref{method:gamma}, with $\adom$ given by
\eqref{MN:asymp.matrix}, and $\areg$ being the matrix of independent
numerical solutions at large $\rho$. In order to calculate them, it is
convenient to rewrite \eqref{MN:eqa} as a system of first-order
equations,\footnote{Notice that both $\mfa$ and $\mfb$ contain 3 components,
  so that \eqref{MN:eq.num} is a system of 6 first order ODEs.
  It is convenient to consider the third component of $\mfa$
  separately, as it fortuitously decouples from the other two. Thus, we have a
  4-component and a 2-component system of first order ODEs.}
\begin{equation}
\label{MN:eq.num}
  \partial_\rho \begin{pmatrix} \mfa \\ \mfb \end{pmatrix} =
  \begin{pmatrix} N & 1 \\ k^2 & M \end{pmatrix} 
  \begin{pmatrix} \mfa \\ \mfb \end{pmatrix}~.
\end{equation}
In order to calculate the three independent regular solutions, we impose
the three initial conditions \eqref{MN:a.reg1}--\eqref{MN:a.reg3}
at very small $\rho$.

\begin{figure}[t]
\begin{center}
 \includegraphics[width=0.6\textwidth]{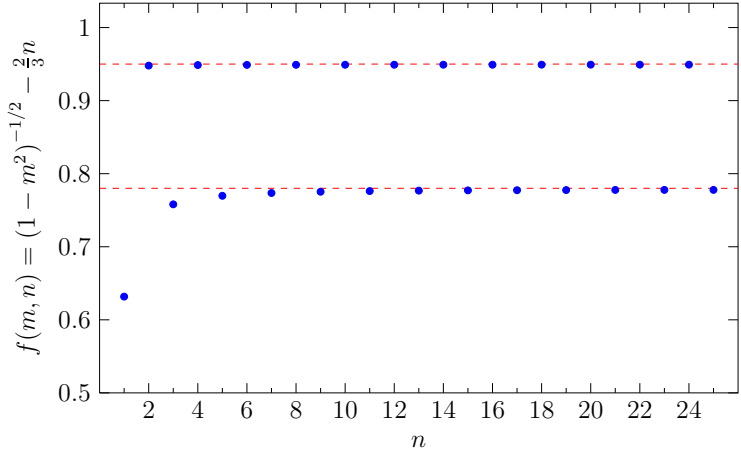}
\end{center}
\vspace{-5mm}
\caption{The function \eqref{MN:fmn} for the numerical mass spectrum
  in the MN background.\label{MN:specplot}}
\end{figure}

Our results are as follows. For the system involving the first two
components of $\mfa$, we obtain discrete mass values. The first few of
these are listed in Tab.~\ref{MN:masstable} and can be compared to the
estimates from the singular background, \eqref{MN:spec3m}. To refine
this comparison, in Fig.~\ref{MN:specplot} we plot  the function 
\begin{equation}
\label{MN:fmn}
  f(m,n) = \left( 1-m^2 \right)^{-1/2} -\frac23 n~,
\end{equation}
which leads to the mass values
\begin{equation}
\label{MN:fmnmass}
  m^2 = 1 - \frac{9}{(2 n + 3 f)^2}~.
\end{equation}
If \eqref{MN:spec3m} were exactly true, $f(m,n)$ would be identical to $1$. We find
that the numerical values consistently come out smaller than those
from the analytical approximation, and that they cluster into two
distinct spectra. For even $n$, $f \approx 0.95$, 
whereas for odd $n$, except for the first few values, $f\approx 0.78$. 
This appearance of two spectra is not too surprising, if we remember
that, in \eqref{MN:spec3m}, values for even and odd $n$ originated from the
first and second component of $\mfa$, respectively. These two components
couple in the regular background so that their spectra will combine into one.
However, the numerical results 
indicate that there are two different kinds of bound states in this spectrum.

%
%

From the massless scalar equation \eqref{MN:a3.eq}, which describes
both the third component of $\mfa$ and the spin-2 field $\mfe^i_j$, we
have not found any discrete mass states, confirming the analytical result of
Sec.~\ref{MN:analytic}.


\section{Klebanov-Strassler system}
\label{KS}

\subsection{KS Background}
\label{KS:bg}

The effective 5-d model describing the bulk dynamics of the KS system
contains seven scalar fields. We will use the Papadopoulos-Tseytlin
\cite{Papadopoulos:2000gj} variables $(x,p,y,\Phi,b,h_1,h_2)$.  
In the following, we shall briefly summarize the general relations for
this system and the KS background solution. For more details, see our
earlier paper \cite{Berg:2005pd} and references therein.

The sigma-model metric is 
\begin{multline}
\label{KS:Gab}
  G_{ab} \partial_\mu \phi^a \partial^\mu \phi^b = 
  \partial_\mu x \partial^\mu x 
  + 6 \partial_\mu p \partial^\mu p
  + \frac12 \partial_\mu y \partial^\mu y
  + \frac14 \partial_\mu \Phi \partial^\mu \Phi 
  + \frac{P^2}2 \e{\Phi-2x} \partial_\mu b \partial^\mu b +\\
  +\frac14 \e{-\Phi-2x} \left[ 
     \e{-2y} \partial_\mu (h_1-h_2) \partial^\mu (h_1-h_2)      
    +\e{2y} \partial_\mu (h_1+h_2) \partial^\mu (h_1+h_2) \right]~,
\end{multline}
and the superpotential reads
\begin{equation}
\label{KS:W}
  W = -\frac12 \left( \e{-2p-2x} +\e{4p} \cosh y \right) 
  +\frac14 \e{4p-2x} \left( Q + 2Pb h_2 + 2Ph_1 \right)~.
\end{equation}
Here, $Q$ and $P$ are constants related to the number of D3-branes and
wrapped D5-branes, respectively.

It is useful to introduce the KS radial variable $\tau$ by 
\begin{equation}
\label{KS:taudef}  
  \partial_\tau = \e{-4p} \partial_r~.
\end{equation}
Here and henceforth, a field variable denotes the
background of that field  ($p$ in this case), while the sigma-model covariant $\mfa$ will be used for
fluctuations. In terms of $\tau$, the KS background solution of
\eqref{method:bg2} is given by 
\begin{align}
\label{KS:Phisol}
  \Phi &= \Phi_0~,\\
\label{KS:ysol}
  \e{y} &= \tanh (\tau/2)~, \\
\label{KS:bsol}
  b &= - \frac{\tau}{\sinh \tau}~, \\
\label{KS:h1sol}
  h_1 &= -\frac{Q}{2P} 
        + P\e{\Phi_0} \coth \tau (\tau\coth \tau-1)~,\\ 
\label{KS:h2sol}
  h_2 &= P\e{\Phi_0} \frac{\tau \coth \tau -1}{\sinh \tau}~,\\
\label{KS:xsol}
  \frac23 \e{6p+2x} &= \coth \tau -\frac{\tau}{\sinh^2 \tau}~,\\
\label{KS:psol}
  \e{2x/3-4p} &= 2 P^2 \e{\Phi_0} 3^{-2/3} h(\tau) \sinh^{4/3} \tau~,
\end{align}
with
\begin{equation}
\label{KS:hsol}
  h(\tau) = \int\limits_\tau^\infty \rmd \vartheta\,
  \frac{\vartheta\coth\vartheta -1}{\sinh^2 \vartheta} \left[
  2\sinh(2\vartheta) -4\vartheta \right]^{1/3}~.
\end{equation}
Moreover, one can show that the warp function $A$ satisfies
\begin{equation}
\label{KS:A}
  \e{-2A-8p} \sim \left(\e{-6p-2x} \sinh \tau \right)^{2/3} h(\tau)~,
\end{equation}
where the proportionality factor depends on an integration constant
that sets the 4-d momentum scale. 


\subsection{KS spin-2 spectrum}
\label{KS:spin2}

\begin{table}[t]
\caption{Mass spectrum of spin-2 glueballs in the KS background ($m^2<100$), 
  and comparison with
  Krasnitz' WKB results \cite{Krasnitz:2003pj}. Krasnitz' values have been
  normalized such that the seventh masses (the highest he calculated)
  agree. \label{KS:Tab1}}
\begin{center}
\begin{tabular}{|r|r|}
\hline
$n$ & Krasnitz \\ \hline
1 &1.06  \\
2 &  2.39  \\
3 &  4.52 \\
4 &  6.65  \\
5 & 9.62  \\
6 &  13.1   \\
7 & 17.1  \\ \hline
\end{tabular}
\hspace{1cm}
\begin{tabular}{|r|r||r|r||r|r|}  \hline
$n$ & $m^2$ &  $n$ & $m^2$ & $n$ & $m^2$ \\  \hline
1 & 1.044 & 
8 & 21.62 & 
15 & 68.78 \\
2 & 2.369 & 
9 & 26.73 & 
16 & 77.69 \\
3 & 4.227 & 
10 & 32.38 &
17 & 87.15 \\
4 & 6.624 &
11 & 38.57 &
18 & 97.15 \\
5 & 9.561 & 
12 & 45.31 &
  &   \\
6 & 13.04 &  
13 & 52.59 &
  &   \\
7 & 17.06 &
14 & 60.41 &
 &  \\    \hline
\end{tabular}
\end{center}
\end{table}

\begin{figure}[t]
\begin{center}
\includegraphics[width=0.5\textwidth]{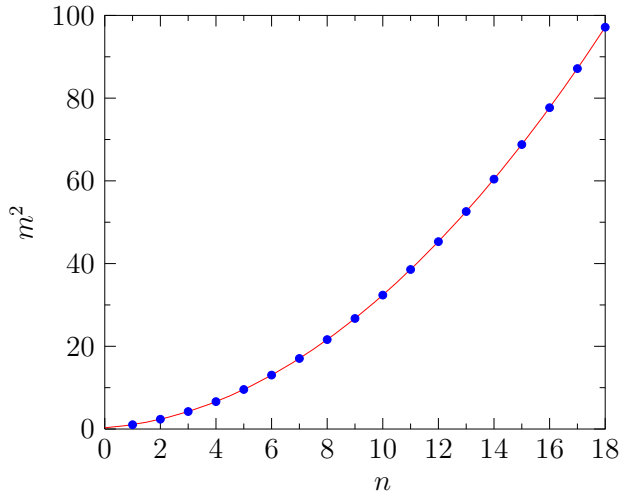}
\vspace{-5mm}
\end{center}
\caption{Mass-squared values of spin-2 states. The solid line
  represents the fit \eqref{KS:spin2fit}.\label{KS:Plot1}}
\end{figure}

Let us start the numerical analysis of fluctuations in the KS
background with the simplest case: the free massless scalar describing
the transverse  traceless components of fluctuations of the bulk
metric.
Specifically, we are interested in the spectrum of spin-2
glueballs, expanding on earlier work by Krasnitz \cite{Krasnitz:2000ir,Krasnitz:2003pj}.

The equation of motion \eqref{method:eqe}, after substituting the
appropriate relations from the previous subsection, takes the form
\begin{equation}
\label{KS:eqe}
  \left( \partial_\tau^2 +2 \e{-6p-2x} \partial_\tau -k^2 \e{-2A-8p}
  \right) \mfe =0~,
\end{equation}
where we omitted tensor indices on $\mfe$. Defining 
\begin{equation}
\label{KS:Idef}
  I(\tau) = \frac{h(\tau)}{h(0)} 
\end{equation}
and appropriately choosing the integration constant in the warp
function \eqref{KS:A} such that
\begin{equation}
\label{KS:Adef}
  \e{-2A-8p} = \left( \e{-6p-2x} \sinh\tau \right)^{2/3} I(\tau)~,
\end{equation}
we can rewrite \eqref{KS:eqe} as
\begin{equation}
\label{KS:eqe.num}
  \left\{ \partial_\tau^2 +2 \e{-6p-2x} \partial_\tau - 
  k^2 I(\tau) \left( \e{-6p-2x} \sinh\tau \right)^{2/3} \right\} \mfe
  =0~. 
\end{equation}
The choice of momentum scale normalization in \eqref{KS:Adef} is a little
unusual, but convenient in what follows.

In this simple one-component system, the method described in
Sec.~\ref{method:method1} is sufficiently robust.\footnote{There is only
  one dominant and one subdominant asymptotic behaviour, and the
  latter is sufficiently suppressed for large $\tau$ with respect to
  the former.}
For this method, first we need to find the regular behaviour at $\tau=0$, from which
the initial conditions will be inferred. For small $\tau$, we find to
leading order $\e{-6p-2x} \approx 1/\tau$ and $I(\tau)\approx 1$, so that
a regular solution of \eqref{KS:eqe.num} is found to behave as 
\begin{equation}
\label{KS:e.regsol}
  \mfe_\text{reg} = 1 + \frac{k^2}6 \tau^2 + \Order{\tau^4}~,
\end{equation}
whereas the singular behaviour, which starts with $1/\tau$, is discarded.
Notice that the choice of momentum scale in \eqref{KS:Adef} makes the
$k^2$ term in \eqref{KS:e.regsol} independent of $h(0)$, a good thing since
the value of $h(0)$ is known only from numerical evaluation of \eqref{KS:hsol}. 

The second step of this method is to consider the asymptotic UV region. For large $\tau$ we have,
again  to leading order, $\e{-6p-2x} \approx 2/3$ and $I(\tau)\sim \tau
\e{-4\tau/3}$. Therefore, the generic dominant asymptotic behaviour 
is just 
\begin{equation}
\label{KS:e.asympt}
  \mfe_\text{dom} = 1 + \Order{\e{-2\tau/3}}~.
\end{equation}
This tells us that we must simply search for values of $k^2$ for which the
regular solution tends to zero for large $\tau$. 

We find a discrete spectrum of spin-2 states, the first of which are
reported and compared to Krasnitz' values in
Table~\ref{KS:Tab1}. Krasnitz' results, which were obtained using a
WKB approximation, are in good agreement with ours. 
As shown in Fig.~\ref{KS:Plot1}, the spectrum fits nicely to a quadratic
curve. A least-square fit yields 
\begin{equation}
\label{KS:spin2fit}
  m^2 \approx 0.2715\, n^2 +0.4936\, n +0.2969~.
\end{equation}
As we will find in the next section, the coefficient of the $n^2$ term
enjoys a certain degree of universality.


\subsection{KS 7-scalar system: Setup}
\label{KS:7scalarsetup}

Now, we turn to the most difficult system of our paper: the seven
coupled scalar fields in the KS background. All scalars appear to be
fully coupled in the bulk, but we follow the insight gleaned from the
``moderate UV'' approximation
\cite{Berg:2005pd} to decouple a $4 \times 4$ set of fields
(the ``glueball sector'') from a $3 \times 3$ set (the
``gluinoball sector'') to leading order in the UV, as will be seen in 
Sec.~\ref{KS:7scalarbcs}. The system of
field equations we consider follows from \eqref{method:eqa} 
upon changing the radial coordinate to $\tau$. One finds
\begin{equation}
\label{KS:7scalar.eom}
 \left[ (\partial_\tau -M) (\partial_\tau-N) -k^2 \e{-2A-8p} \right]
 \mfa =0~,
\end{equation}
where we have dropped the tensor indices, and the matrices $M$ and $N$
are defined by 
\begin{equation}
\begin{split}
\label{KS:7scalar.MNdef}
  M^a_{~b} &= -N^a_{~b} -K^a_{~b} -2 \e{-2x-6p} \delta^a_b~,\\
  N^a_{~b} &= \e{-4p} \left( \partial_b W^a -\frac{W^aW_b}{W} \right)~,\\
  K^a_{~b} &= 2\e{-4p} \G{a}{bc}W^c~.
\end{split}
\end{equation}
As before, we fix the momentum scale as in \eqref{KS:Adef}.

Now, the matrices $M$ and $N$ \emph{a priori} depend on the constants $P$ and
$\Phi_0$. However, there is a linear field
transformation that removes this dependence from \eqref{KS:7scalar.eom}. 
This implies that these constants can affect the mass spectrum 
only by an overall change of the momentum scale, which is not 
visible in our effective 5-d approach.
Starting with the fluctuation vector $\mfa=
(\delta x,\delta p,\delta y,\delta\Phi,\delta b,\delta h_1,\delta h_2)^T$, 
the linear transformation that accomplishes this is $\mfa' = R \mfa$ with
\begin{equation}
\label{KS:rotateda}
 \mfa'=\left(\delta x, \delta p, \frac{\delta h_1}{P\e{\Phi_0}}, 
          \delta \Phi, \delta y, \delta b, \frac{\delta
          h_2}{Pe^{\Phi_0}} \right)^T
\end{equation}
and we also rotate the matrices by, \eg $N' = R N R^{-1}$. 
Henceforth, we shall consider the rotated matrices,
dropping primes. 
The somewhat lengthy expressions for the rotated matrices $K$ and
$N$ (that are now independent of $P$ and $\Phi_0$ as advertised) as well as the sigma-model
metric $G$ can be found in Appendix~\ref{appendix:KSmats}.


\subsection{KS 7-scalar system: Boundary conditions}
\label{KS:7scalarbcs}

In this section, we will consider the asymptotic (large-$\tau$) and
deep-IR (small-$\tau$)
behaviour of the solutions of \eqref{KS:7scalar.eom},
which are needed to fix the boundary (initial) conditions for the
numerical integration. 

Let us start with the large-$\tau$ region. Asymptotically, the matrices and the warp
term  in \eqref{KS:7scalar.eom} can be expanded in
powers of $\e{-\tau}$, such that\footnote{The ``coefficients'' may contain rational
  functions of $\tau$, but no exponentials.}
\begin{equation}
\label{KS:KNexp}
  K = K^{(0)} + \e{-\tau} K^{(1)} + \Order{\e{-2 \tau}}~, \quad 
  N = N^{(0)} + \e{-\tau} N^{(1)} + \Order{\e{-2 \tau}}~.
\end{equation}
In what follows, we can always drop the $\Order{\e{-2\tau}}$ terms (the reason 
for this will become clearer later on, cf.\ the discussion below (\ref{KS:7scalar.asyeqn2})).
For the sake of brevity, we write the matrices in block form, 
\begin{equation}
\label{KS:matrixblock}
  K = \begin{pmatrix} 
    K_{4\times 4} & K_{4 \times 3} \\
    K_{3\times 4} & K_{3 \times 3} 
    \end{pmatrix}~,
\end{equation}
and analogously for $N$. Then, the matrices in \eqref{KS:KNexp} are
\begin{align}
\label{KS:K044}
  K^{(0)}_{4\times 4} &= 
  \begin{pmatrix}
    0 & 0 & \frac{2}{3(\tau-1/4)} & 0 \\
    0 & 0 & 0 & 0 \\
    -2& 0 & -\frac{1}{\tau-1/4} & -1 \\
    0 & 0 & \frac{4}{3(\tau-1/4)} & 0 
  \end{pmatrix}~, \\
\label{KS:K033}
  K^{(0)}_{3\times 3} &= 
  \begin{pmatrix}
    0 & 0 & - \frac{4}{3(\tau-1/4)} \\
    0 & -\frac{1}{\tau-1/4} & 0 \\
    2 & 0 & -\frac{1}{\tau-1/4} 
  \end{pmatrix}~, \\
\label{KS:K034} 
  K^{(0)}_{4\times 3} &= K^{(0)}_{3\times 4} = 0~,
\end{align}
\begin{align}
\label{KS:N044}
  N^{(0)}_{4\times 4} &= 
  \begin{pmatrix}
    -\frac{1}{\tau+1/4} & -\frac{4\tau-1}{\tau+1/4} &
    -\frac{2}{3(\tau+1/4)} & 0 \\
    -\frac{2(\tau-1/4)}{3(\tau+1/4)} &   
    -\frac{2(\tau+5/4)}{3(\tau+1/4)} & \frac{2}{9(\tau+1/4)} & 0 \\
    \frac{1}{\tau+1/4} & \frac{4\tau-1}{\tau+1/4} &
    \frac{2}{3(\tau+1/4)} & 1 \\
    0 & 0 & 0 & 0 
  \end{pmatrix}~,\\
\label{KS:N033}
  N^{(0)}_{3\times 3} &= 
  \begin{pmatrix}
     -1 & 0 & 0 \\
      0 & 0 & 1 \\
     -2 & 1 & 0 
  \end{pmatrix}~,\\
\label{KS:N034} 
  N^{(0)}_{4\times 3} &= N^{(0)}_{3\times 4} = 0~,
\end{align}
and
\begin{align}
\label{KS:K143}
  K^{(1)}_{4\times 3} &= 
  \begin{pmatrix}
    0 & \frac{4(\tau-1)}{3(\tau-1/4)} & 
    -\frac{4\tau}{3(\tau-1/4)} \\
    0 & 0 & 0 \\
    -4(\tau-2) & 0 & 4 \\
    0 & -\frac{8(\tau-1)}{3(\tau-1/4)} & -\frac{8\tau}{3(\tau-1/4)} 
  \end{pmatrix}~, \\
\label{KS:K134}
  K^{(1)}_{3\times 4} &= 
  \begin{pmatrix}
    0 & 0 & \frac{8\tau}{3(\tau-1/4)} & 0 \\
    -4(\tau-1) & 0 & 0 & 2(\tau-1) \\
     4(\tau-2) & 0 & 4 & 2(\tau-2) 
  \end{pmatrix}~, \\
\label{KS:K133}
  K^{(1)}_{4\times 4} &= K^{(1)}_{3\times 3} = 0~,
\end{align}
\begin{align}
\label{KS:N143}
  N^{(1)}_{4\times 3} &=
  \begin{pmatrix}
    \frac{1}{\tau+1/4} & -\frac{4(\tau-1)}{3(\tau+1/4)} &
    \frac{4\tau}{3(\tau+1/4)} \\ 
    \frac{2(\tau-1/4)}{3(\tau+1/4)} & \frac{4(\tau-1)}{9(\tau+1/4)} &
    -\frac{4\tau}{9(\tau+1/4)} \\
    \frac{4\tau^2 -5\tau-5/2}{\tau+1/4} & \frac{16\tau-1}{3(\tau+1/4)}
    &  -\frac{4\tau}{3(\tau+1/4)} \\
    0 & 0 & 0 
  \end{pmatrix}~, \\
\label{KS:N134}
  N^{(1)}_{3\times 4} &=
  \begin{pmatrix}
    \frac{2}{\tau+1/4} & \frac{8(\tau-1/4)}{\tau+1/4} &    
    \frac{4}{3(\tau+1/4)} & 0 \\
    \frac{2(\tau-1)}{\tau+1/4} & \frac{8(\tau-1)(\tau-1/4)}{\tau+1/4}
    & \frac{4(\tau-1)}{3(\tau+1/4)} & -2(\tau-1) \\
    -\frac{2(\tau-2)}{\tau+1/4} &
    -\frac{8(\tau-2)(\tau-1/4)}{\tau+1/4} &
    -\frac{4(\tau-2)}{3(\tau+1/4)} & -2(\tau-2) 
  \end{pmatrix}~,\\
\label{KS:N133}
  N^{(1)}_{4\times 4} &= N^{(1)}_{3\times 3} = 0~.
\end{align}
The transformation $\mfa \rightarrow \mfa'$, cf.\ \eqref{KS:rotateda}, 
has brought the matrices into this nice block form. 
We also need
\begin{equation}
\label{KS:xpasy}
  \e{-2x-6p} = \frac23 +\Order{\e{-2\tau}}
 \end{equation}
as well as 
\begin{equation}
\label{KS:warpinf}
  \e{-2A-8p} = \frac{3^{1/3}}{h(0)} \left(\tau-\frac14\right) \e{-2\tau/3} 
  \left[ 1 +\Order{\e{-2\tau}} \right]~.
\end{equation}
It is a useful check that the leading-order terms of these
expressions coincide with the respective quantities evaluated in the Klebanov-Tseytlin 
background \cite{Klebanov:2000nc}. 

Since \eqref{KS:warpinf} is exponentially suppressed, the leading
order terms of the asymptotic solutions will be independent of the
momentum $k$. 
We note that this is just as we assumed in Sec.~\ref{method:polestruc}. 
In contrast, in the MN system, the asymptotic behaviours of the
solutions depend on $k$ [cf.\ \eqref{MN:a2}].

The asymptotic UV solutions, including the leading and some subleading
terms, are found by iteratively solving the equations
\begin{align}
\label{KS:7scalar.asyeqn}
  \left( \partial_\tau-N^{(0)} \right) \phi^{(n)} &= \psi^{(n)} 
   + \e{-\tau} N^{(1)} \phi^{(n-1)}~, \\
\label{KS:7scalar.asyeqn2}
  \left( \partial_\tau-M^{(0)} \right) \psi^{(n)} &= 
  \beta \left(\tau-\frac14\right) \e{-2 \tau/3} \phi^{(n-1)} 
   + \e{-\tau} M^{(1)} \psi^{(n-1)}~,
\end{align}
where  $\beta = 3^{1/3}k^2 /h(0)$, and we set $\phi^{(-1)}=\psi^{(-1)}=0$. The solutions $\phi^{(0)}$ are
the leading order terms of the asymptotic solutions.  
Again, we will drop all $\Order{\e{-2\tau}}$ terms in the iteration. The reason for this is the following. 
Dropping subleading terms in the initial conditions leads to
systematic errors in the numerical solutions, which must obviously 
be kept smaller than the relevant parts of the solutions. This implies
that the initial conditions of all behaviours we use must be given to
the same order in $\e{-\tau}$. We will use the mid-point method of
Sec.~\ref{method:method2}, in which only the subdominant asymptotic
behaviours are needed. Now, the leading term of the weakest subdominant
behaviour goes like $\e{-8\tau/3}$, whereas the strongest subdominant
behaviour goes like $\e{-\tau}$. Thus, they differ by $\e{-5\tau/3}$
and we can drop $\Order{\e{-2\tau}}$
corrections.\footnote{The same argument holds for the method of
  Sec.~\ref{method:method1}, where only the dominant solutions are
  needed. The weakest dominant solution goes like $\e{-\tau/3}$, the
  strongest one like $\e{4\tau/3}$.}

The somewhat lengthy expressions of the asymptotic solutions are
deferred to Appendix~\ref{KSasympt}.

Now, let us consider the small-$\tau$ region. In order to find the
independent behaviours for small $\tau$, we expand the matrices
in \eqref{KS:7scalar.eom} about $\tau=0$ and make an ansatz of
the form 
\begin{equation}
\label{KS:7scalar.regansatz}
  \mfa(\tau) = \tau^q \left( \mfa_0 + \tau \mfa_1 + \tau^2 \mfa_2 +
  \tau^3 \mfa_3 +\cdots \right)~,
\end{equation}
with some undetermined number $q$, and constant vectors $\mfa_n$,
$n=0,1,2,\ldots$. These are then determined recursively from the
equation of motion (using computer algebra). 
It turns out that, of the 14 solutions, 
there exist four solutions with $q=0$, one with $q=1$, three with
$q=2$, four with $q=-1$, and one each with $q=-2$ and $q=-3$. Thus,
there exist eight solutions, whose components are finite at $\tau=0$,
and six with singular component behaviour. 
However, the regularity condition \eqref{method:regcond} tells us that one
of the solutions with $q=0$ is, in fact, singular. 
Therefore, we arrive at precisely seven regular and seven singular
solutions. The independent small-$\tau$ behaviours are listed 
in Appendix~\ref{app:KSregsols}.


\subsection{KS spin-0 spectrum}
\label{KS:spin0spec}

Analogously to the MN case, instead of 7 coupled 2nd order ODEs,
it is more convenient to solve  14 coupled 1st order ODEs.
Thus, let us rewrite \eqref{KS:7scalar.eom} as
\begin{equation}
\label{KS:eq.num}
  \partial_\tau \begin{pmatrix} \mfa \\ \mfb \end{pmatrix} =
  \begin{pmatrix} N & 1 \\ k^2 \e{-2A-8p} & M \end{pmatrix} 
  \begin{pmatrix} \mfa \\ \mfb \end{pmatrix}~.
\end{equation}

\begin{figure}[ht]
\begin{center}
\includegraphics[width=0.8\textwidth]{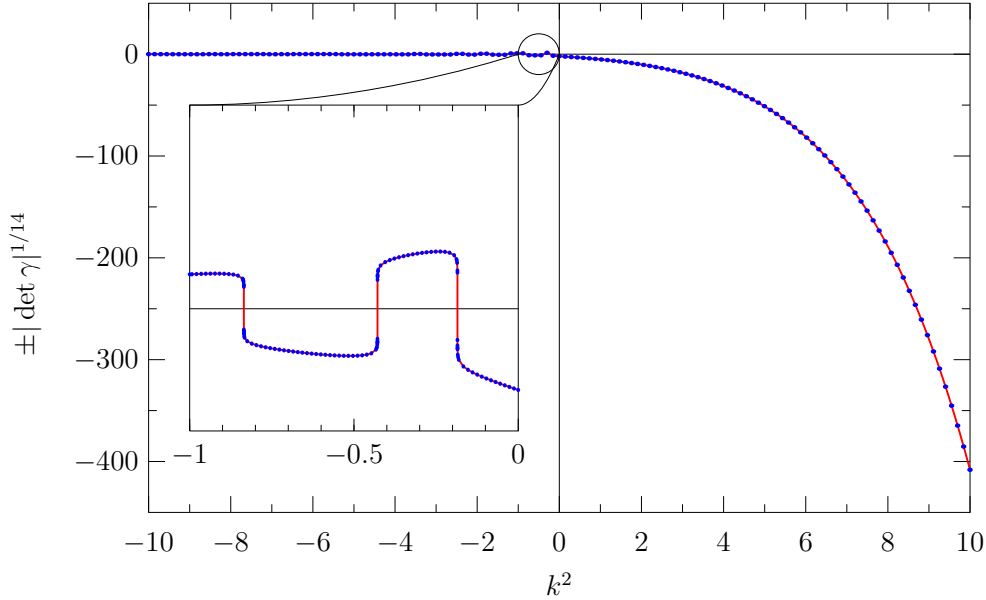}
\caption{$\det \gamma$ as a function of $k^2$. For clarity, we have 
taken the 14th root of the absolute value of the determinant (leaving
its sign untouched). The plot shows 
clearly that there are no zero crossings for positive $k^2$. The inset 
shows the zeros of $\det \gamma$ for the first three spin-0 states.
\label{detplots}}
\end{center}
\end{figure}

\begin{figure}[ht]
\begin{center}
\includegraphics[width=0.5\textwidth]{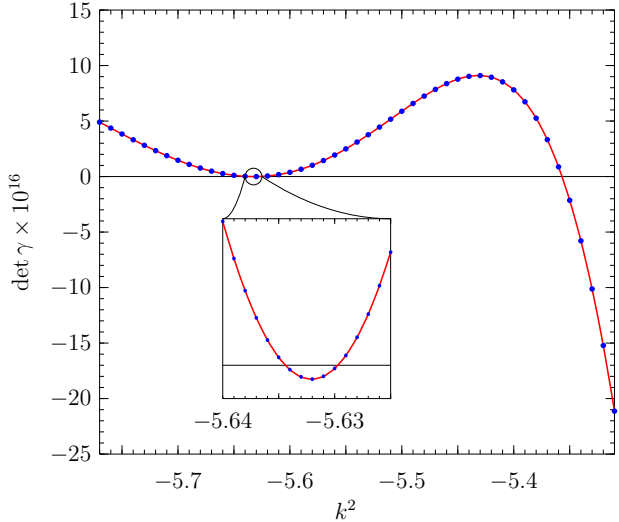}
\caption{Here we plot $\det \gamma$ in the range $k^2 \in [-5.8 \ldots -5.3]$. 
The inset zooms in on the region around $k^2=-5.63$, where there are two zeros.
\label{KS:detsmooth2}}
\end{center}
\vspace{-5mm}
\end{figure}

\begin{table}[h]
\begin{center}
\begin{tabular}{|c|c|c|c|c|c|c|c|} \hline 
$m^2$ & 0.185 & 0.428 & 0.835 & 1.28 & 1.63 & 1.94 & 2.34    \\  \hline
      & 2.61 & 3.32 & 3.54 & 4.12 & 4.18 & 4.43 & 4.43     \\ \hline 
     &  5.35 & 5.63 & 5.63 & 6.59 & 6.66 & 6.77 & 7.14   \\ \hline 
\end{tabular}
\end{center}
\vspace{-5mm}
\caption{Low-lying Klebanov-Strassler spin-0 mass states,
extracted from zero-crossings in Fig.~\ref{detplots}.
More values are given in Appendix \ref{KSspec}.}
\label{KSmasstable}
\end{table}

We apply the midpoint determinant method
as outlined in section \ref{method:method2}. That is, we compute
 the $14\times 14$ matrix 
\begin{equation}
\label{KS:mid-point}
\gamma=
  \begin{pmatrix} \areg & \asub \\
      \breg & \bsub
      \end{pmatrix}_{\tau=\tau_{\rm mid}} 
\end{equation}
as a function of $k^2$ and look for zero crossings of $\det\gamma$.
A rough plot of $\det \gamma$ is shown in Fig.~\ref{detplots}. We verify that 
there are no zero crossings for positive $k^2$. The zero crossings of 
$\det \gamma$ can be found by zooming in on a particular region of the 
plot, as shown, for example, in the inset. 
As the determinant itself changes over many orders of magnitude, it is useful to plot 
 $\pm |\det \gamma|^{1/14}  \equiv
\mathrm{sign}(\det \gamma) |\det \gamma|^{1/14}$ instead.
The sharp turns in the inset are merely
artifacts of this, as is apparent from Fig.~\ref{KS:detsmooth2}, 
where $\det \gamma$ itself is displayed without the 14th root, and there are no sharp turns. 

We have calculated all mass values up to $m^2=600$. The first few are listed in Table~\ref{KSmasstable}; for a  more extended list see Table~\ref{Table:KSextended} in Appendix \ref{KSspec}. Our results exhibit two interesting features. First, it appears that several mass values are nearly degenerate, such as the two values about $m^2\approx 4.43$ and two values about $m^2\approx 5.63$. The origin of this in our calculation can be seen in Fig.~\ref{KS:detsmooth2}: The $\det \gamma$ curve barely seems to touch the $k^2$-axis, and only zooming in reveals that there are two crossings. As we do not have a dynamical explanation of it, it is possible that this near degeneracy is accidental, or it may indicate a multiplet originating from a weakly broken symmetry of the low-energy effective theory. 

\begin{figure}[ht]
\begin{center}  
  \includegraphics[width=0.43\textwidth]{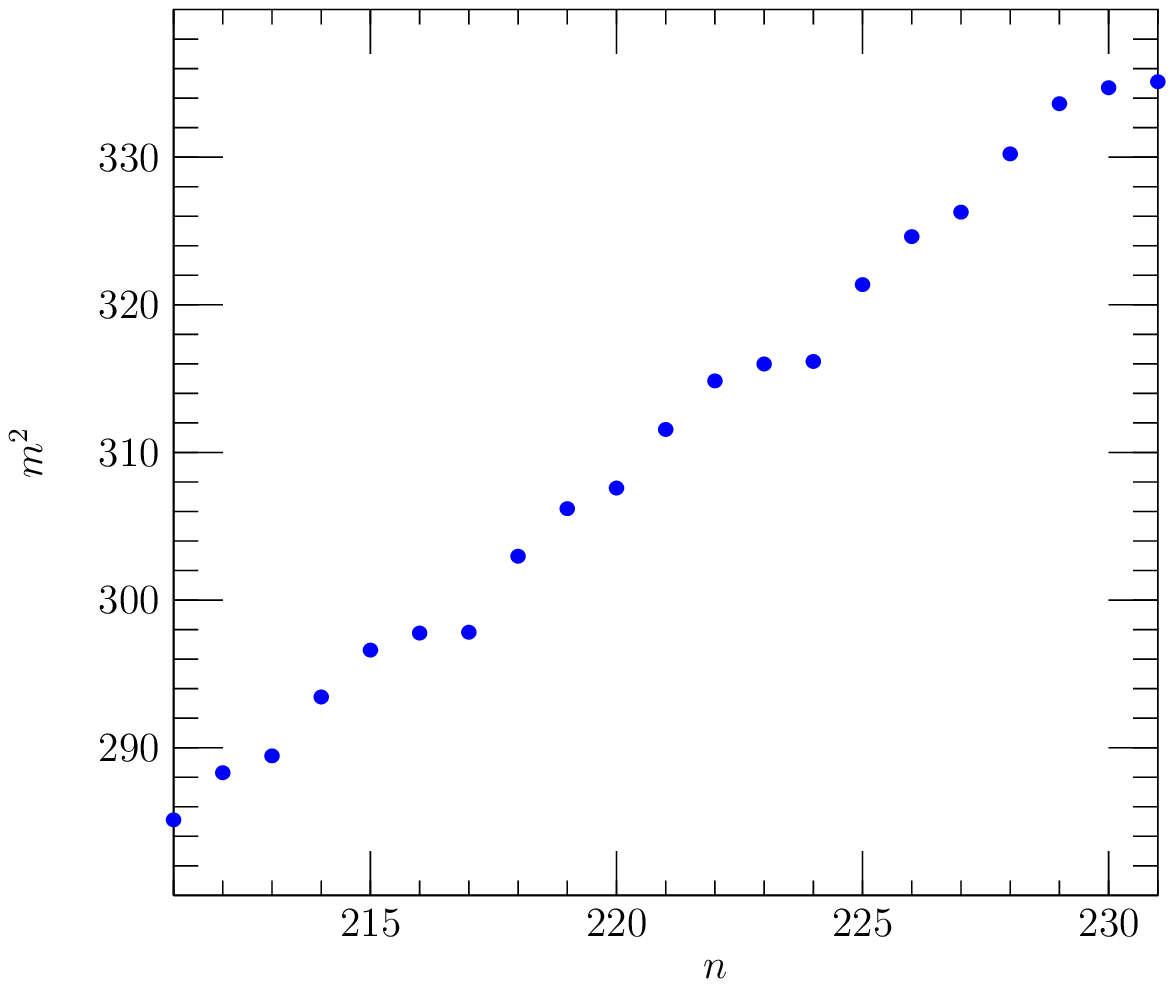}
\hspace{0.5cm}
  \includegraphics[width=0.4\textwidth]{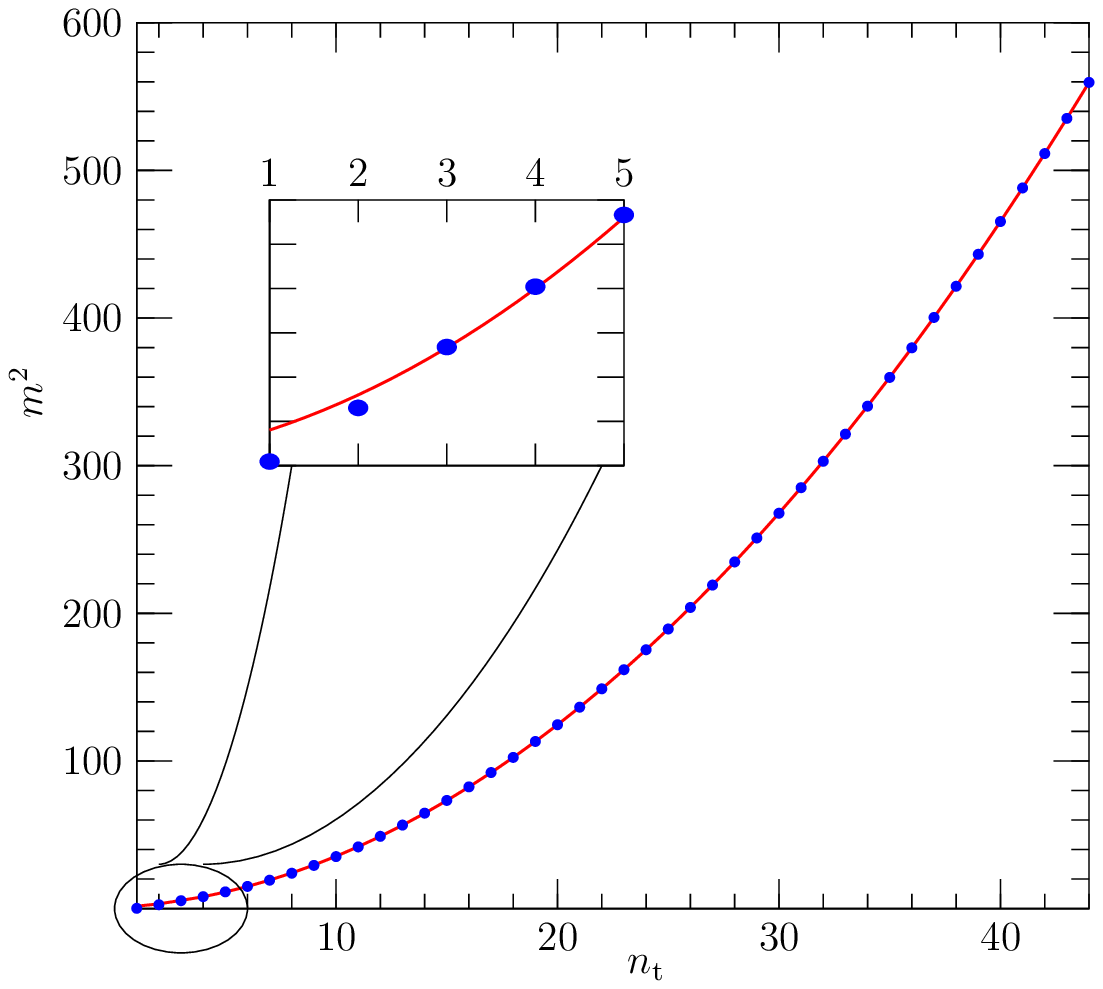}
\caption{On the left, we see that for high $m^2$, it is easy to
  identify the periodicity ``empirically'' as 7. On the right, 
we have extracted one tower
  (\ie every 7th point) from the full spectrum, enumerated states in that tower by 
a parameter $n_t$ (the ``excitation number'') and fit those points 
to a quadratic curve starting with the third point.
As shown in the inset,  $m^2 \sim n_t^2$ 
is not a good description for very low $n_t$, unlike in the spin-2 case.
\label{towerplots}}
\end{center}
\end{figure}

The second interesting feature is the appearance, for large masses, of a periodic pattern of period 7 in the consecutive spectrum excitation number $n$. On the left hand side of Fig.~\ref{towerplots}, this pattern is easily visible. Hence, we split the spectrum into 7 towers of mass states, one of which is shown on the right hand side of Fig.~\ref{towerplots}. From now on, we 
distinguish excitation numbers of individual towers, denoted by $n_t$,
from the excitation number $n$ of the whole spin-0 spectrum.

Like the spin-2 spectrum, the seven spin-0 spectra also exhibit 
quadratic dependence of $m^2$ on excitation number $n_t$.
Leaving out the first two values of each tower (i.e.\ $n_t=1,2$) least square fits 
of the  ``large $m^2$'' behavior  
(using mass values up to 
$m^2=600$) yield
\begin{equation}
\label{KS:spin0fit}
\begin{split}
  m^2 & \approx  0.271\, n_t^2 +0.774\, n_t +0.562~, \\
  m^2 & \approx  0.270\, n_t^2 +0.928\, n_t -0.430~, \\
  m^2 & \approx  0.275\, n_t^2 +0.769\, n_t +1.921~, \\
  m^2 & \approx  0.272\, n_t^2 +1.017\, n_t +1.023~, \\
  m^2 & \approx  0.272\, n_t^2 +1.119\, n_t +0.337~, \\
  m^2 & \approx  0.271\, n_t^2 +1.150\, n_t +0.648~, \\
  m^2 & \approx  0.273\, n_t^2 +1.082\, n_t +2.544~.
\end{split}
\end{equation}
The last digit of the leading coefficients (i.e.\ the coefficients of the 
$n_t^2$ terms) can vary slightly if one performs the fits with 
more than two states from each tower left out. Still, it is obvious that  
within some uncertainties the leading coefficients all coincide rather 
well, both amongst each other and also with the coefficient of the spin-2 tower \eqref{KS:spin2fit}, 
leading to roughly universal behaviour at large $n_t$.


\subsection{KS 7-scalar system: Quadratic confinement}
\label{sec:mixing}

If one plots the {\it experimental} values for squared meson
masses $m^2$ (e.g.\ for the first few resonances of the $\rho$ meson) 
against their consecutive number $n$, the data 
obeys $m^2 \sim n$ to good accuracy.\footnote{Since
there should be no confusion here, we omit the subscript ``t" on $n_t$ from now on.}
One might call this ``linear confinement''.
Karch, Katz, Son and Stephanov (KKSS)
\cite{Karch:2006pv} emphasized that there are no known supergravity
backgrounds that reproduce this experimental fact about mesons
from AdS/CFT.

In fact, the claim of KKSS was even stronger (p.\ 13): that
for all known supergravity backgrounds $m^2 \sim n^2$,
which is ``quadratic confinement''.
Extending these arguments to our setting,
one might expect to see quadratic confinement in, say, the
Klebanov-Strassler background. A priori, given the complexity of fluctuations around the
Klebanov-Strassler solution, 
one could have expected that no such general statement could
be made except possibly in some extreme asymptotic regime.

This expectation turns out to be wrong.
Although we cannot be absolutely sure from our analysis, the
KKSS claim that $m^2 \sim n^2$
seems resoundingly apparent in Fig.~\ref{towerplots} and the fits 
\eqref{KS:spin0fit}. 
Since the KS theory is in any case not QCD, linear confinement is not crucial to have, 
but our results strengthen the case for finding controllable
models that do exhibit linear confinement  (the metric-scalar model given
in \cite{Karch:2006pv} seems promising, but so far has not been shown to be the solution
of any known theory).

Of course, one would expect that much of the low-energy dynamics of the
theory would be determined by the lowest-lying mass states, and they
have a richer structure than the large-$n$ asymptotics --- a glimpse of this can be seen 
in the inset on the right hand side of Fig.~\ref{towerplots}. 
Since the KS theory incorporates spontaneous breaking of chiral symmetry,  
this should have left its mark on our spectrum. We do observe some interesting patterns, 
but we have not been able to link them conclusively to symmetry breaking. This deserves further study.

As alluded to in the introduction, operator mixing is even more subtle
here than in standard AdS/CFT. We have
computed the spin-0 mass spectrum, but we have not said what the mass
states are composed of in terms of dual gauge theory operators. Some of the problems of 
identifying the mass states were already addressed in section 
\ref{method:polestruc}, i.e.\ the ambiguity in the 
normalization of the dominant asymtptotic solutions and the 
freedom to add to a given dominant solution multiples of asymptotic
solutions of equal or weaker strength. Thus we do not attempt
to give a further interpretation of the mass states. 
We just content ourselves by noticing that there is actually a quantity 
that is easy to obtain with our method, namely the coefficients
$d_{\lambda,i}$ from (\ref{method:eigen.subdom}). These can 
be determined straightforwardly by solving (\ref{method:mid-point})
at the various mass values. This gives information on the composition of 
the ``cavity mode'' (that we called $\mfa_\lambda$ in chapter \ref{method}) 
corresponding to a mass state in terms of 
our basis of subdominant solutions. However, as the interpretation 
in terms of operator mixing is not clear to us, we refrain from 
determining these coefficients explicitly here. Moreover, also the quantities
$Z_{\lambda,i}$, defined in (\ref{method:Zdef2}), could be determined in principle. 
They correspond to the residues of the mass poles if $N_\lambda$, introduced in 
(\ref{method:Ydef}), is really one, as 
in the case of asymptotic AdS spaces (see Appendix \ref{appads}).

\subsection{Comparison to hardwall approximations}
\label{sec:hardwall}

\begin{table}[t]
\begin{center}
\begin{tabular}{|c||c|c|c|} \hline 
values from Table \ref{KSmasstable} & $\tau_{\rm IR}=0.5$ & $\tau_{\rm IR}=0.75$ 
& $\tau_{\rm IR}=1$  \\  
\hline 
0.19 & 0.27 & 0.30 & 0.34 \\
0.42 & 0.56 & 0.63 & 0.71 \\
0.83 & 0.86 & 1.03 & 1.14 \\
1.29 & 1.39 & 1.62 & 1.83 \\
1.63 & 1.78 & 1.89 & 2.03 \\
1.94 & 2.32 & 2.42 & 2.57 \\
2.35 & 2.43 & 2.56 & 2.75 \\
2.61 & 3.26 & 3.39 & 3.70 \\
3.33 & 3.58 & 3.80 & 4.09 \\
3.53 & 4.06 & 4.22 & 4.40 \\
4.13 & 4.30 & 4.60 & \\
4.17 & 4.54 & 4.81 & \\
4.43 (2x) & 4.71 & &\\
 \hline
\end{tabular}
\end{center}
\vspace{-5mm}
\caption{Comparison of our mass values with a toy 
hardwall model as described in the text (only values up to $m^2=5$
are included). This 
demonstrates that the spectrum indeed 
strongly depends on the IR dynamics.}
\label{hardwalltable}
\end{table}

In this section, we would like to give some arguments supporting our conviction
stated in the introduction that the detailed form 
of the spectrum depends crucially on the details of the IR dynamics in 
the gauge theory. Holographically, this means that it is important to use 
the full Klebanov-Strassler background to determine the low 
lying mass eigenvalues. To illustrate this, we compare our spectrum with 
a naive toy hardwall model. In particular, one might hope to find the right spectrum 
by using the Klebanov-Tseytlin approximation of the KS background. 
This is roughly equivalent to cutting off the KS background at some value 
$\tau_{\rm IR}$ of order one (the two solutions differ strongly only for
$\tau \le 0.8$). In order to use the determinant method described in 
Sec.~\ref{method:method1}, one would like to impose seven independent 
initial conditions at $\tau_{\rm IR}$ and then solve the equations numerically
from the IR towards 
the UV. One again demands that there is a linear combination 
of those numerical solutions  that 
does not contain any dominant solutions $\mfa_{\rm dom}$ when expanded in a basis of asymptotic 
solutions. 
Lacking any guiding principle for choosing these initial conditions, 
we arbitrarily impose that all components of the seven scalar 
fields should vanish at $\tau_{\rm IR}$, whereas for each initial condition 
one of the first derivatives are taken to be unity. This gives seven 
independent initial conditions. The resulting spectrum is contained 
in Table~\ref{hardwalltable} for several values of $\tau_{\rm IR}$, 
together with the correct mass values. Obviously, the ``toy spectra'' 
depend on the value of $\tau_{\rm IR}$, but they also differ strongly from 
the correct values. We believe that this holds more
generally, even if one had a different set of initial conditions
than these, whether or not they are physically motivated.

To summarize, we do not see any way of implementing a hard-wall approximation that reproduces the correct spectrum of Table~\ref{KSmasstable}.

\section{Outlook}
\label{outlook}

We have computed spin-0 and spin-2 mass spectra for the
Maldacena-Nunez (wrapped D5-brane) and Klebanov-Strassler (warped
deformed conifold) backgrounds. Although fairly complicated, there are
some simple features. For example, for large excitation number $n$, the spin-0 states 
in the KS theory organize into 7 towers, each of which
displays quadratic confinement, in agreement with the claims 
of \cite{Karch:2006pv}. (``Large'' is not terribly large;
around $n\geq 3$.) For low excitation number $n$, there is a rich
structure of mass values, which exhibits interesting patterns of near degeneracy.

For the Maldacena-Nunez background, our numerical results confirm the rough features of
the analytic spectra we derived in \cite{Berg:2005pd} and improved 
upon here; the discrete mass spectrum has an upper bound, which is quite different from the KS case.

The situation with respect to other studies of glueball spectra in
these backgrounds, based on various truncations, has not changed from \cite{Berg:2005pd}.
The one result we have been able to compare our results to is the set of
mass states for spin-2 glueballs in KS, where the first few values were
computed in the WKB approximation by Krasnitz in \cite{Krasnitz:2003pj},
and are roughly correct, as in Table~\ref{KS:Tab1}.

It would be very important to understand the details of mixing in this context; without
further information along the lines we have suggested, we cannot determine the composition of the mass 
states we find.
On the good side, the spectrum we have found is unique, in that the variables used to find it are 
gauge-invariant, and we impose the boundary conditions one really wants to impose, rather than 
approximations thereof.

These results and methods could straightforwardly be applied to
other gauge/gravity dual pairs. Apart from other $\N=1$ dualities,
it would be interesting to apply these methods also in 
$\N=0$ gauge theory. There, the existence of a simple superpotential $W$,
which greatly simplified our task, is not guaranteed but occurs e.g.\ for
$\N=0$
backgrounds with ``pseudo-Killing'' spinors (cf.\ for example
\cite{Skenderis:2006jq}). Here is a concrete question: do the
meson spectra computed in other backgrounds contain
glueball admixtures from the $e^{-\phi}$ in the DBI action? 

Independently of supersymmetry breaking, including probe
branes in Klebanov-Strassler \emph{\`a la} 
\cite{Sakai:2003wu, Kuperstein:2004hy} would be interesting to study using
our methods.

Finally, the greatest challenge for holography in presently known confining backgrounds 
is to compute correlators from the Klebanov-Strassler background.
Some progress was made in
\cite{Krasnitz:2000ir,Krasnitz:2002ct,Krasnitz:2003pj,Aharony:2005zr,Aharony:2006ce}. We 
considered the Krasnitz (extreme high-energy) limit in \cite{Berg:2005pd} and found some analytical solutions.
However, those solutions were checked to be valid around $k^2 \sim 10^6$, whereas in this paper, 
we were interested in IR physics, and thus the Krasnitz approximation was not available to us. In 
fact, we only considered solutions up to $k^2 \sim 2000$, and going much higher presents a numerical 
difficulty, as one needs to go increasingly far in the UV for the solutions to become asymptotic, so 
that the determinant method is applicable. This is a superable challenge for connecting the high-energy 
and low-energy regimes, which might 
ultimately be the most useful aspect of holography for confining gauge theories. We hope to return to 
these issues in future publications.

\section*{Acknowledgements}

It is a pleasure to thank Massimo Bianchi, Andreas Karch, Emanuel
Katz, Albion Lawrence, Scott Noble,
Carlos Nu\~nez, Henning Samtleben, David Tong and Amos Yarom for helpful discussions and comments. 
This work is supported in part by the European Community's Human Potential Program under 
contract MRTN-CT-2004-005104 'Constituents, fundamental forces and symmetries of the 
universe'. This research was supported in part by the National Science Foundation
under Grant No. PHY99-07949.
The work of M.~B. is supported by European Community's Human Potential
Program under contract MRTN-CT-2004-512194, `The European Superstring
Theory Network'. 
He would like to thank the 
Galileo Galilei Institute in Florence for hospitality.
The work of M.\ H.\ is supported by the German Research Foundation (DFG) within 
the Emmy Noether-Program (grant number: HA 3448/3-1). 
Both M.~B. and M.~H would like to thank the KITP
in Santa Barbara for hospitality during the 
program ``String Phenomenology''.
The work of W.\ M.\ is supported in part by the Italian Ministry of Education and Research (MIUR), project 2005-023102.


\begin{appendix}
\section{Some 2-point functions in AdS/CFT}
\label{appads}

In this section, we review 
how the 2-point functions for some known cases of AdS/CFT
arise as sums over the spectrum of bulk eigenfunctions,
using expressions developed in the main text. 

\subsection{AdS bulk} 

Let us start with a set of $n$ free massive scalar fields on
$(d+1)$-dimensional AdS bulk space. 
Using the language of Sec.~\ref{method}, we will derive the 2-point
function as a ``sum'' over the spectrum of bulk eigenfunctions. (For
pure AdS the spectrum is continuous, so the ``sum'' is in
fact an integral. This will be different in the next subsection.)
When the smoke has cleared  (in \eqref{appads:regsol}),
the connection to the standard computation
in terms of the IR-regular Bessel function $\rmK_{\nu}(kz)$ will
be evident.

The bulk equation of motion \eqref{method:eqa} (in $d$-momentum space) is
\begin{equation}
\label{appads:eom}
  \left( \partial_z^2 -\frac{d-1}{z} \partial_z -\frac{m^2}{z^2} -k^2 \right) \phi = 0~,  
\end{equation}
where $z= \e{-r}$ is the radial coordinate. We have omitted the field
indices, and $m^2$ is to be understood as a diagonal matrix with
entries $m_i^2$, $i=1,\ldots,n$. The conformal dimensions of the operators $\op_i$
dual to the components of $\phi$ are related to the mass parameters
$m_i$ by\footnote{In general one needs to consider
 $\Delta^{\pm}_i = \frac{d}2 \pm \alpha_i$, but for the purposes
of this discussion we restrict to the upper sign. For simplicity, we
also consider generic (non-integer) values of $\alpha_i$. }
\begin{equation}
\label{appads:dim}
  \Delta_i = \frac{d}2 +\alpha_i~,\qquad \alpha_i = \sqrt{\frac{d^2}4+m_i^2}~.
\end{equation}
Small values of $z$ give the asymptotic UV region, whereas large $z$
describe the bulk interior (IR region). 
 
Conventionally normalized asymptotic solutions of \eqref{appads:eom} are\footnote{We have reinstated the field index $a$.}
\begin{align}
\label{appads:asympsols1}
  \phidom_i^a(z,k) &= \delta^a_i \Gamma(1-\alpha_i) \left( \frac{k}{2} \right)^{\alpha_i} 
     z^{d/2} \rmI_{-\alpha_i} (kz) = \delta^a_i \, z^{d/2-\alpha_i} +\cdots~, \\ 
\label{appads:asympsols2} 
  \phisub_i^a(z,k) &= \delta^a_i \Gamma(1+\alpha_i) \left( \frac{k}{2} \right)^{-\alpha_i}  
     z^{d/2} \rmI_{\alpha_i} (kz) = \delta^a_i \, z^{d/2+\alpha_i} +\cdots~,
\end{align}
where the $\rmI_\alpha$ are modified Bessel functions, and after the second equal signs we indicated just the respective leading behaviours. The powers of $k$ in front of the solutions are necessary in order to make the leading terms $k$-independent, while the remaining coefficients are conventional normalizations. 

For asymptotically AdS bulk backgrounds, \eqref{appads:eom} still holds in the asymptotic region. Therefore,  
the leading behaviours of the asymptotic solutions remain as in \eqref{appads:asympsols1} and \eqref{appads:asympsols2}, 
\begin{equation}
\label{appads:asympgen}
   \phidom_i^a(z,k) = \delta^a_i \, z^{d/2-\alpha_i}+\cdots~,\qquad
   \phisub_i^a(z,k) = \delta^a_i \, z^{d/2+\alpha_i}+\cdots~.
\end{equation}
The general form of the matrix $Z_{ij}$ \eqref{method:Zdef} in asymptotically AdS spaces follows easily. One finds
\begin{equation}
\label{appads:Zgen}
   Z_{ij} = 2 \alpha_i\, \delta_{ij}~.
\end{equation}

A general result of holographic renormalization \cite{Bianchi:2001kw, Martelli:2002sp, Papadimitriou:2004ap}
is that (the non-local part of) the exact one-point function $\vev{\op_i}_\text{exact}$ is 
\begin{equation}
\label{appads:exact1pt}
  \vev{\op_i}_\text{exact} = 2 \alpha_i \, d_i~  
\stackrel{\eqref{appads:dim}}{=} \left(2\Delta_i-d\right)d_i                 \; .
\end{equation}
Comparing this with \eqref{method:exact1pt}, one finds
\begin{equation}
\label{appads:Yij}
  Y_{ij} = 2 \alpha_i \,\delta_{ij} = Z_{ij}~, 
\end{equation}
which, together with \eqref{method:Ydef}, implies $N_\lambda=1$.

Returning to the pure AdS bulk, \eqref{appads:eom} admits a continuous spectrum of regular 
and subdominant solutions for $k^2=-\lambda^2$, $\lambda>0$. The eigenfunctions are given 
by\footnote{The generic label $\lambda$ for the eigenfunctions used in the main text 
[c.f.\ \eqref{method:green.ansatz}] is replaced here by two indices, $\lambda=-k^2$ 
and $i=1,\ldots,n$. As before, the upper index $a$ is the vector component index.} 
\begin{equation}
\label{appads:eigenfun}
  \phi_{\lambda i}^a(z) = \delta^a_i \, \sqrt{\lambda}\,z^{d/2} \rmJ_{\alpha_i} (\lambda z)
\end{equation}
and satisfy the orthogonality relation
\begin{equation}
\label{appads:ortho}
  \int\limits_0^\infty \frac{\rmd z}{z} z^{-(d-2)} \phi_{\lambda i}(z) \cdot \phi_{\lambda' j}(z) =
     \delta(\lambda-\lambda') \delta_{ij}~.
\end{equation}
%
%
%
%
%
Considering the small-$z$ behaviour of \eqref{appads:eigenfun}, one can read off the response coefficients $d_{\lambda i,j}$, 
\begin{equation}
\label{appads:dlambda}
  d_{\lambda i,j} = \delta_{ij} \left(\frac{\lambda}{2} \right)^{\alpha_i} \frac{\sqrt{\lambda}}{\Gamma(1+\alpha_i)}~.
\end{equation}
Hence, after using \eqref{method:Zdef2} and $N_\lambda=1$, one obtains for \eqref{method:2ptfin} 
\begin{equation}
\label{appads:2ptads}
  \vev{\op_i(k)\op_j(-k)} = 
    \delta_{ij} \frac{2^{2(1-\alpha_i)}}{\Gamma(\alpha_i)^2} 
    \int\limits_0^\infty \rmd \lambda\, \frac{\lambda^{2\alpha_i+1}}{k^2+\lambda^2} = 
    -\delta_{ij}\, 2 \frac{\Gamma(1-\alpha_i)}{\Gamma(\alpha_i)} \left( \frac{k}2 \right)^{2\alpha_i}~.
\end{equation}
Notice that the second equality holds only after analytic
continuation, because the integral does not exist if
$\alpha_i\geq0$. This is equivalent to adding an infinite contact term
to the integral over the spectrum. For example, if $0<\alpha_i<1$, we
rewrite the integrand as
\begin{equation}
\label{appads:rewriteinteg}
  \frac{\lambda^{2\alpha_i+1}}{k^2+\lambda^2} = \lambda^{2\alpha_i-1} 
  - k^2 \frac{\lambda^{2\alpha_i-1}}{k^2+\lambda^2}
\end{equation}
and add a counter term that cancels the first term on the right hand side. 

The finite result \eqref{appads:2ptads} can also be obtained in the usual way, \ie by considering a regular solution of \eqref{appads:eom},
\begin{equation}
\label{appads:regsol}
   \phi_i^a(z,k) = \delta_i^a \frac{2 \tilde{d}}{\Gamma(\alpha_i)}
   \left(\frac{k}{2}\right)^{\alpha_i} z^{d/2} \rmK_{\alpha_i}(kz)~
=\delta_i^a \tilde{d}\, \frac{2^{1-\alpha_i}}{\Gamma(\alpha_i)}  
   \int\limits_0^\infty \rmd \lambda\, \frac{\lambda^{\alpha_i+1}}{k^2+\lambda^2} 
     z^{d/2} \rmJ_{\alpha_i}(\lambda z)  \; ,   
\end{equation}   
reading off the relation between the response and source coefficients, $d$ and $\tilde{d}$, with 
the help of 
\begin{equation}
\label{}
 \rmK_{\alpha}(x) 
= 2^{\alpha -1} \Gamma (\alpha) x^{-\alpha} (1+ \ldots) - 
 2^{-\alpha-1} \frac{\Gamma(1-\alpha)}{\alpha} x^\alpha (1+ \ldots)
\end{equation}
and then using \eqref{appads:exact1pt}. It is interesting to note that the small-$z$ behaviour 
of the integrand in \eqref{appads:regsol} is only subdominant, while
the
expression in terms of $\rmK_{\alpha_i}$ contains 
also a dominant piece. The explanation of this apparent discrepancy is
the UV part of the 
spectrum (large $\lambda$), \ie one cannot find some sufficiently small $z$ such that 
$\lambda z$ is also small for all values of $\lambda$. Hence, loosely, 
the UV-part of the spectrum generates the dominant behaviour. 

\subsection{Active scalar in GPPZ flow}

As a second example, let us take the active scalar in the GPPZ flow 
\cite{Girardello:1999bd}. 
Its gauge invariant bulk equation of 
motion reads \cite{Bianchi:2003ug, Muck:2004qg}
\begin{equation}
\label{appGPPZ:eom}
  \left[ u(1-u) \partial_u^2 +(3u-2) \partial_u +\frac{3u}{4(1-u)} -1 +\frac{k^2}4 \right] \phi =0~, 
\end{equation}
where the radial coordinate $u$ is defined by $u=1-\e{-2r}$, and the warp factor is $\e{2A} = u/(1-u)$. 

Eq.~\eqref{appGPPZ:eom} admits a discrete spectrum of regular and subdominant eigenfunctions, with mass squares
\begin{equation}
\label{appGPPZ:spectrum}
  m_n^2= 4n(n+1)~,\qquad n=1,2,3,\ldots\ .
\end{equation}
The normalized eigenfunctions are 
\begin{equation}
\label{appGPPZ:eigenfun}
  \phi_n = \sqrt{\frac{2(2n+1)}{n(n+1)}} (1-u)^{3/2} \frac{\rmd}{\rmd u} \rmP_n(2u-1)~,
\end{equation}
where $\rmP_n$ are Legendre polynomials.\footnote{As in \cite{Bianchi:2003ug, Muck:2004qg}, regular and subdominant solutions of \eqref{appGPPZ:eom} are given by Jacobi polynomials $\rmP_{n-1}^{(1,1)}(z)$, which are proportional to $\rmd \rmP_n(z)/ \rmd z$.} 
One easily finds the response coefficients
\begin{equation}
\label{appGPPZ:dn}
  d_n = \sqrt{2n(n+1)(2n+1)}~.
\end{equation}
Thus, using $\alpha=1$ (we have a $\Delta=3$ operator) and \eqref{appads:Yij}
(which holds for asymptotically AdS spaces), we obtain for the 2-point function \eqref{method:2ptfin} 
\begin{equation}
\label{appGPPZ:2pt}
  \vev{\op(k) \op(-k)} = \sum\limits_{n=1}^\infty \frac{8n(n+1)(2n+1)}{k^2 +4n(n+1)} 
  +\text{c.t.}\ .  
\end{equation}
Clearly, the sum in \eqref{appGPPZ:2pt} does not converge, so that there are again infinite contact terms. It is instructive to compare \eqref{appGPPZ:2pt} with the finite result from holographic renormalization \cite{Bianchi:2001de, Muck:2001cy},
\begin{equation}
\label{appGPPZ:2ptfinite}
  \vev{\op(k) \op(-k)} = \frac{k^2}2 \left[ 
  \psi\left(\frac{3+\sqrt{1-k^2}}2 \right) + \psi\left(\frac{3-\sqrt{1-k^2}}2 \right) 
  -\psi(1) -\psi(2) \right]~,
\end{equation}
where $\psi(z) = [\ln \Gamma(z)]'$. Using the identity
\begin{equation}
\label{appGPPZ:ident}
  \psi(x) -\psi(y) = \sum\limits_{n=0}^\infty \left( \frac1{y+n} -\frac1{x+n} \right)~,    
\end{equation}
we obtain from \eqref{appGPPZ:2ptfinite}
\begin{equation}
\label{appGPPZ:2ptsum}
\begin{split}
  \vev{\op(k) \op(-k)}
  &= \frac{k^4}2 \sum\limits_{n=1}^\infty \frac{2n+1}{n(n+1)[k^2 +4n(n+1)]} \\
  &= \sum\limits_{n=1}^\infty \left[ \frac{8n(n+1)(2n+1)}{k^2 +4n(n+1)} 
    -2(2n+1) +k^2 \frac{2n+1}{2n(n+1)} \right]~.
\end{split}
\end{equation}
Again, we see that holographic renormalization makes a precise choice for the contact terms and, therefore, picks a renormalization scheme and scale. Indeed, the finite result \eqref{appGPPZ:2ptfinite} satisfies the renormalization conditions
\begin{equation}
\label{appGPPZ:2ptrencond}
  \left. \vev{\op(k) \op(-k)} \right|_{k^2=0} = 0~, \qquad
  \frac{\rmd}{\rmd k^2} \left. \vev{\op(k) \op(-k)} \right|_{k^2=0} = 0~.
\end{equation}

\section{Matrices for the MN background}
\label{app:MNmats}

The entries of the matrices $M$ and $N$ of
Eq.~\eqref{MN:eqa} for the regular MN background ($c=0$) are 
\begin{equation}
\label{app:MN.M}
\begin{split}
M_{11}&= -[ (8\rho^2-4\rho+1) +
         \e{-4\rho}(-64\rho^3+112\rho^2-48\rho+2) +\\
&\qquad +
	 \e{-8\rho}(256\rho^4-896\rho^3+248\rho^2+108\rho-17) +\\
&\qquad +
	 \e{-12\rho}(1536\rho^4-736\rho^2+28) + \\
&\qquad + 
	 \e{-16\rho}(256\rho^4+896\rho^3+248\rho^2-108\rho-17) +\\
&\qquad +
	 \e{-20\rho}(64\rho^3+112\rho^2+48\rho+2) +
	 \e{-24\rho}(8\rho^2+4\rho+1) ] / \\
&\quad / \{ \rho (1-\e{-4\rho})^4 [1-\e{-4\rho}(4\rho+1)] 
         (4\rho-1+\e{-4\rho}) \} \\
M_{12} &= [ (3\rho-1) +
         \e{-4\rho}(16\rho^3-48\rho^2+12\rho+2) +
	 \e{-8\rho}(96\rho^3-30\rho) +\\
&\qquad
	 +\e{-12\rho}(16\rho^3+48\rho^2+12\rho-2) +
	 \e{-16\rho}(3 \rho+1) ] \times \\
&\quad  \times 4 \e{-2\rho} / \{ \rho (1-\e{-4\rho})^3 [1-\e{-4\rho}(4\rho+1)] (4\rho-1+\e{-4\rho}) \} \\
M_{21}&= -[1-\e{-4\rho}(4\rho+1)] (4\rho-1+\e{-4\rho}) [(\rho-1) +
         \e{-4\rho} 6\rho + \e{-8\rho} (\rho+1) ] \times \\
&\quad \times 4\e{-2\rho} /
         [ \rho (1-\e{-4\rho})^5] \\
M_{22}&= -[ (2 \rho-1) +
         \e{-4\rho}(32\rho^3-128\rho^2+56\rho-2) +\\
&\qquad +
	 \e{-8\rho}(-256\rho^4+960\rho^3-256\rho^2-118\rho+17) + \\
&\qquad +
	 \e{-12\rho}(-1536\rho^4+768\rho^2-28) + \\
&\qquad +
	 \e{-16\rho}(-256\rho^4-960\rho^3-256\rho^2+118\rho+17) +\\
&\qquad +
	 \e{-20\rho}(-32\rho^3-128\rho^2-56\rho-2) +
	 \e{-24\rho} (-2\rho-1) ] / \\
&\quad / \{ \rho (1-\e{-4\rho})^4 [1-\e{-4\rho}(4\rho+1)] (4\rho-1+\e{-4\rho}) \} \\
M_{33}&= -8\rho (1-\e{-4\rho})^2 / \{ [1-\e{-4\rho}(4\rho+1)] (4\rho-1+\e{-4\rho}) \} \\
M_{13}&=M_{23}=M_{31}=M_{32}=0~,
\end{split}
\end{equation}
and
\begin{equation}
\label{app:MN.N}
\begin{split}
N_{11}&= -[ 1 +
         \e{-4\rho}(16\rho^2-32\rho+4) +
	 \e{-8\rho}(96\rho^2-10) +
	 \e{-12\rho}(16\rho^2+32\rho+4) +\\
&\qquad +
	 \e{-16\rho} ] / [ \rho (1-\e{-4\rho})^4 ] \\
N_{12}&= [(\rho-1) + \e{-4\rho} 6\rho + \e{-8\rho} (\rho+1) ] 
         \times 4\e{-2\rho} / [ \rho (1-\e{-4\rho})^3 ] \\
N_{21}&= - [ (3\rho-1) + 
         \e{-4\rho}(16\rho^3-48\rho^2+12\rho+2) +
	 \e{-8\rho}(96\rho^3-30\rho) +\\
&\qquad +
	 \e{-12\rho}(16\rho^3+48\rho^2+12\rho-2) +
	 \e{-16\rho}(3\rho+1) ] \times 
         4 \e{-2\rho} / [ \rho (1-\e{-4\rho})^5 ] \\
N_{22}&= - [ (2\rho-1) +
         \e{-4\rho}(-16\rho^2+28\rho-4) +
	 \e{-8\rho}(-96\rho^2+10) +\\
&\qquad +
	 \e{-12\rho}(-16\rho^2-28\rho-4) +
	 \e{-16\rho}(-2\rho-1) ] / [ \rho (1-\e{-4\rho})^4 ] \\
N_{13}&=N_{23}=N_{31}=N_{32}=N_{33}=0~.
\end{split}
\end{equation}

\section{Bulky KS stuff}
%
%
\subsection{KS matrices}
\label{appendix:KSmats}

We present here the explicit expressions for the $7\times 7$ matrices
appearing in Section~\ref{KS:7scalarsetup}. To shorten the
formulae, a number of abbreviations will be used. First,
\begin{equation}
\label{appKS:csdef}
  c = \cosh y = \coth \tau~,\qquad 
  s = \sinh y = -(\sinh \tau)^{-1}~,
\end{equation}
where $y$ denotes the background field of Section~\ref{KS:bg}. 
Second, we introduce 
\begin{equation}
\label{appKS:b12def}
  B_1 = \tau c -1~, \qquad 
  B_2 = \tau s^2 -c ~,
\end{equation}
and 
\begin{align}
\label{appKS:A1def}
  A_1 &= h(\tau) \left( 4 s B_2 \right)^{-1/3} 
      = h(\tau) \sinh \tau \left( 2\sinh 2\tau -4\tau \right)^{-1/3}~,\\
\label{appKS:A2def}
  A_2 &= -A_1 \left( c B_2 -\frac23 \right) -\frac12 s B_1 B_2~.
\end{align}

Let us consider the behaviour for small and large $\tau$
of $A_1$ and $A_2$. As $h(0)$ is a finite, positive constant, one
obtains
\begin{equation}
\label{appKS:A12.smalltau}
  A_1(0) = \frac12 3^{1/3} h(0)~, \qquad
  A_2(0) = \frac43 A_1(0)~.
\end{equation}
For large $\tau$, starting from 
\begin{equation}
\label{appKS:h.largetau}
  h(\tau) \approx 3 \e{-4\tau/3} \left(\tau-\frac14 \right)~,
\end{equation}
one obtains
\begin{equation}
\label{appKS:A12.largetau}
  A_1(\tau) \approx \frac32 \e{-\tau} \left(\tau-\frac14
  \right)~,\qquad
  A_2(\tau) \approx \frac32 \e{-\tau} \left(\tau+\frac14
  \right)~.  
\end{equation}

With the abbreviations \eqref{appKS:csdef}--\eqref{appKS:A2def}, 
the (rotated) matrices 
$K^a_{~b}= 2 \e{-4p} \G{a}{bc} W^c$ and $N^a_{~b}$ are given by
\begin{equation}
\label{appKS:K}
  K = \begin{pmatrix}
    0 & 0 & \frac{s}{2A_1 B_2} & 0 & 
    0 & -\frac{s^2 B_1}{2A_1 B_2} & \frac{s^2 \tau}{2A_1 B_2} \\
    0 & 0 & 0 & 0 & 0 & 0 & 0 \\
    2(1+2cB_2) & 0 & \frac{2(A_2+cA_1B_2)}{A_1 B_2} & 2c B_2 +1 &
    2s( \tau+2B_2) & 0 & -2s \\
    0 & 0 & \frac{s}{A_1 B_2} & 0 & 
    0 & \frac{s^2 B_1}{A_1 B_2} & \frac{s^2 \tau}{A_1 B_2} \\
    0 & 0 & -\frac{s^2 \tau}{A_1 B_2} & 0 &
    0 & 0 & -\frac{s}{A_1B_2} \\
    2sB_1 & 0 & 0 & -s B_1 &
    0 & \frac{2(A_2+cA_1B_2)}{A_1 B_2} & 0 \\
    -2s (\tau+2B_2) & 0 & -2s & -s (\tau+2B_2) &
    -2(1+2cB_2) & 0 &  \frac{2(A_2+cA_1B_2)}{A_1 B_2} 
      \end{pmatrix}~,
\end{equation}
\begin{multline}
\label{appKS:N}
  N = \left( \begin{matrix}
    -\frac{2c(A_2+c A_1 B_2)}{A_2} & -\frac{4cA_1}{A_2} & 
    \frac{cs}{2A_2} & 0 \\
    -\frac{2cA_1}{3A_2} & -\frac{s B_1+ 2c A_1}{A_2} & 
    \frac{s}{6A_2 B_2} & 0 \\
    \frac{-2(1+2cB_2)(A_2+cA_1B_2)}{A_2} & 
    -\frac{4A_1(1+2cB_2)}{A_2} & \frac{s(1+2cB_2)}{2A_2} & -(1+2cB_2) \\
    0 & 0 & 0 & 0 \\
    - \frac{2s(A_2+cA_1B_2)}{A_2} & -\frac{4sA_1}{A_2} & 
    \frac{s^2}{2A_2} & 0 \\
    -\frac{2sB_1(A_2+cA_1B_2)}{A_2} & -\frac{4sA_1B_1}{A_2} &
    \frac{s^2B_1}{2A_2} & sB_1 \\
    \frac{2s(\tau+2B_2)(A_2+cA_1B_2)}{A_2} &
    \frac{4sA_1(\tau+2B_2)}{A_2} & -\frac{s^2(\tau+2B_2)}{2A_2} &
    s(\tau+2B_2) 
    \end{matrix} \right. \\
   \left. \begin{matrix} 
    -\frac{s(A_2+cA_1B_2)}{A_2} & 
    -\frac{c s^2 B_1}{2A_2} & \frac{c s^2 \tau}{2A_2} \\
    -\frac{sA_1}{3A_2} & -\frac{s^2 B_1}{6A_2 B_2} &
    \frac{s^2 \tau}{6A_2B_2} \\ 
    -\frac{2sA_2(\tau+2B_2) +sA_1B_2(1+2cB_2)}{A_2} & 
    -\frac{s^2B_1(1+2cB_2)+4csA_2}{2A_2} & 
    \frac{s^2 \tau(1+2cB_2)}{2A_2} \\
    0 & 0 & 0 \\
    -\frac{s^2 A_1B_2}{A_2} -c & - \frac{s^3 B_1}{2A_2} &
    \frac{s^3 \tau}{2A_2} \\
    -\frac{s^2 A_1 B_1 B_2}{A_2} & 
    -\frac{s^3 B_1^2}{2A_2} & 
    \frac{s^3 \tau B_1}{2A_2} +1  \\
    \frac{2A_2(1+2cB_2) +s^2 A_1 B_2(\tau+2B_2)}{A_2} &
    \frac{s^3 B_1(\tau+2B_2) +2A_2(2s^2+1)}{2A_2} &
    - \frac{s^3\tau (\tau+2B_2)}{2A_2}
  \end{matrix}\right)~,
\end{multline}
The matrix $M$ is given by
\begin{equation}
\label{appKS:M}
  M = -N -K +\frac{4}{3B_2} \mathbb{I}~,
\end{equation}
where $\mathbb{I}$ denotes the $7\times 7$ unit matrix.

Finally, we also need the sigma-model metric for the rotated
fluctuation fields. It transforms as $G' = (R^{-1})^T G 
R^{-1}$, where $R$ is the linear transformation matrix that leads to
\eqref{KS:rotateda}, and the superscript $T$ denotes the
transpose. Explicitly, we find 
\begin{equation}
\label{KS:7scalar.Grot}
  G' = 
  \begin{pmatrix}
    1 & 0 & 0 & 0 & 0 & 0 & 0 \\
    0 & 6 & 0 & 0 & 0 & 0 & 0 \\
    0 & 0 & \frac12 P^2\e{\Phi_0-2x} \cosh(2y) & 0 & 0 & 0 & 
      \frac12 P^2\e{\Phi_0-2x} \sinh(2y) \\
    0 & 0 & 0 & \frac14 & 0 & 0 & 0 \\
    0 & 0 & 0 & 0 & \frac12 & 0 & 0 \\
    0 & 0 & 0 & 0 & 0 & \frac12 P^2\e{\Phi_0-2x} & 0 \\
    0 & 0 & \frac12 P^2\e{\Phi_0-2x} \sinh(2y) & 0 & 0 & 0 & 
      \frac12 P^2\e{\Phi_0-2x} \cosh(2y)
  \end{pmatrix}~,
\end{equation}
where for $x$ and $y$ one should substitute the respective 
backgound solutions.

\subsection{KS 7-scalar system: Asymptotic solutions}
\label{KSasympt}

In this appendix, we list as reference the asymptotic solutions of KS the
7-scalar system. They are found by iteratively solving 
\eqref{KS:7scalar.asyeqn} and \eqref{KS:7scalar.asyeqn2}.

For convenience, we abbreviate
\begin{equation}
  \beta = \frac{3^{1/3}}{h(0)} k^2~.
\end{equation}
Moreover, $\bzero_3$ and $\bzero_4$ denote 3 and 4 zero vector
entries, respectively.

The dominant asymptotic (large $\tau$) solutions, up to and including terms of order
$\e{-\tau/3}$, are 
\begin{multline}
\label{KSasympt:domsol1}
\frac{\e{4\tau/3}}{4 \tau +1} \begin{pmatrix}
-12 \\ 4 \\ 12 \\ 0 \\ \bzero_3
\end{pmatrix}
+ \frac{9\beta}{32(4\tau+1)} \e{2\tau/3} \begin{pmatrix}
6(5+4\tau) \\ - (9+4\tau) \\ - 6(5+4\tau)
 \\ 0 \\ \bzero_3
\end{pmatrix}
+ \frac{24}{4\tau+1} \e{\tau/3} \begin{pmatrix}
\bzero_4 \\ 1 \\ \tau -1 \\ 2-\tau  
\end{pmatrix} \\
+ \frac{27\beta^2}{256(4\tau+1)} \begin{pmatrix}
- 24\tau^2-48\tau-63/2 \\ 8\tau+9 \\ 
24\tau^2+48\tau+63/2 \\ 0 \\ \bzero_3
\end{pmatrix}
-\frac{27\beta (4\tau+5)}{8(4\tau+1)} e^{-\tau/3} \begin{pmatrix} 
\bzero_4 \\ 1 \\ \tau-1 \\ 2-\tau
\end{pmatrix}~,
\end{multline}
\begin{equation}
\label{KSasympt:domsol2}
\e{\tau} \begin{pmatrix}
\bzero_4 \\ 0 \\ 1 \\ 1 
\end{pmatrix}
+ \frac{9\beta}{32} \e{\tau/3} \begin{pmatrix}
\bzero_4 \\ 2 \\ 2-2\tau \\ -1-2\tau
\end{pmatrix}
+ \begin{pmatrix}
2 \\ -2/3 \\ 4\tau-2 \\ 0 \\ \bzero_3
\end{pmatrix} 
- \frac{9\beta^2}{256} \e{-\tau/3} \begin{pmatrix}
\bzero_4 \\ 8 \tau^2 - 30\tau + 45 \\ -6\tau^2 - 39\tau + 243/2 \\ 
6\tau^2+6\tau-81
\end{pmatrix}~,
\end{equation}
\begin{equation}
\label{KSasympt:domsol3}
\frac{\e{2\tau/3}}{4 \tau +1} \begin{pmatrix}
4\tau+13 \\ 2\tau-7/2 \\ 12\tau-9 \\ 0 \\ \bzero_3
\end{pmatrix}
- \frac{\beta}{32} \begin{pmatrix}
36\tau+63 \\ 8 \tau-18 \\ 24\tau^2+12\tau-279/2 \\ 72\tau+42 
\\ \bzero_3
\end{pmatrix}
-\frac{e^{-\tau/3}}{4 \tau +1} 
\begin{pmatrix} 
\bzero_4 \\ 64\tau^2-104\tau-6 \\
120\tau^2-246\tau-99 \\ -24\tau^2+174\tau+99
\end{pmatrix}~,
\end{equation}
\begin{equation}
\label{KSasympt:domsol4}
\begin{pmatrix}
1/2 \\ -1/6\\ \tau-1 \\ 1 \\ \bzero_3
\end{pmatrix}~,
\end{equation}
\begin{equation}
\label{KSasympt:domsol5}
\frac{4}{4\tau+1}
\begin{pmatrix}
1/2 \\ -1/6 \\ \tau-1/4\\ 0 \\ \bzero_3
\end{pmatrix}~,
\end{equation}
\begin{equation}
\label{KSasympt:domsol6}
\e{-\tau/3}
\begin{pmatrix}
\bzero_4 \\ 2\tau+1 \\ 3\tau+3/2 \\ 9/4 
\end{pmatrix}
\end{equation}
and
\begin{equation}
\label{KSasympt:domsol7}
\e{-\tau/3}
\begin{pmatrix}
\bzero_4 \\ 4 \\ 9 \\ -3
\end{pmatrix}~.
\end{equation}

The subdominant asymptotic (large $\tau$) solutions, up to and including terms of order
$\e{-8\tau/3}$ (why exactly $\e{-8\tau/3}$ is explained in section
\ref{KS:7scalarbcs}), are 
\begin{multline}
\label{KSasympt:subsol1}
\e{-\tau} \begin{pmatrix}
\bzero_4 \\ 1 \\ \tau \\ 1-\tau 
\end{pmatrix}
+ \frac{9\beta}{16} \e{-5\tau/3} 
\begin{pmatrix}
\bzero_4 \\ 4\tau+11 \\ 2\tau^2 +\tau-12 \\ -2\tau^2+\tau+85/4
\end{pmatrix}
+ \frac{\e{-2\tau}}{20} 
\begin{pmatrix}
32\tau-29 \\ \frac16(16\tau+37) \\ -7(16\tau-7) \\ 0 \\ \bzero_3
\end{pmatrix} \\
+ \frac{27\beta^2}{6400} \e{-7\tau/3} 
\begin{pmatrix}
\bzero_4 \\ 300 \tau^2 +1625 \tau+627 \\
80 \tau^3 -33 \tau^2 -1131\tau -13649/10\\
-\frac{3}{40} (1600\tau^3 +2440\tau^2 -35920\tau-27383)
\end{pmatrix} \\
+ \frac{\beta}{16000(4\tau+1)} e^{-8\tau/3} 
\begin{pmatrix} 
3(35200\tau^3 -27760 \tau^2-316608 \tau+42739) \\ 
4(800\tau^3+ 4240\tau^2 -15228\tau-33961)\\ 
-3(169600\tau^3+ 219680\tau^2 -705996\tau -67973) \\ 0 \\ \bzero_3
\end{pmatrix}~,
\end{multline}
\begin{multline}
\label{KSasympt:subsol2}
\e{-\tau} \begin{pmatrix}
\bzero_4 \\ 0 \\ 1 \\ -1 
\end{pmatrix}
+ \frac{9\beta}{8} \e{-5\tau/3} 
\begin{pmatrix}
\bzero_4 \\ 1 \\ \tau+1/8 \\ -\tau+5/8
\end{pmatrix}
+ \e{-2\tau}
\begin{pmatrix}
-3 \\ -13/6 \\ 1 \\ 0 \\ \bzero_3
\end{pmatrix} \\
+ \frac{3\beta^2}{640} \e{-7\tau/3} 
\begin{pmatrix}
\bzero_4 \\ 195\tau +2989/12 \\
\frac1{50}(3150 \tau^2 +1505 \tau -5591) \\
-\frac1{100}(11700 \tau^2 -5260\tau-28993)
\end{pmatrix} \\
+ \frac{3\beta}{400(4\tau+1)} e^{-8\tau/3} 
\begin{pmatrix} 
-3(40\tau^2+682\tau+69) \\ 
-(4\tau+11/2)(140\tau+57) \\ 
-3(620\tau^2 -709\tau -117) \\ 0 \\ \bzero_3
\end{pmatrix}~,
\end{multline}
\begin{multline}
\label{KSasympt:subsol3}
\e{-4\tau/3} \begin{pmatrix}
1 \\ -1 \\ 6\tau-3 \\ -4\tau+9 \\ \bzero_3
\end{pmatrix}
+ \frac{3\beta}{800} \e{-2\tau} 
\begin{pmatrix}
-80\tau^2-32\tau-291 \\ -40\tau^2-356\tau-5807/6 \\ 
1680\tau^2+2732\tau-1634 \\ -50(4\tau-15)(4\tau+9) \\ 
\bzero_3
\end{pmatrix} \\
+ \frac{3}{250} \e{-7\tau/3}
\begin{pmatrix}
\bzero_4 \\
2000\tau/3 \\ -400\tau^2 +260\tau-547 \\ -400\tau^2+2260\tau-297 
\end{pmatrix} \\
+ \frac{\beta^2}{204800} \e{-8\tau/3} 
\begin{pmatrix}
 -34560\tau^3 -111024\tau^2 +259556\tau-497927/2 \\ 
 -11520\tau^3 -104688\tau^2 -1489028\tau/3 -2917225/6 \\ 
 470880\tau^3 +1839672\tau^2 -135995\tau -766688 \\ 
-3(57600\tau^3-36000\tau^2-1110600\tau-1043879) \\ 
\bzero_3
\end{pmatrix}~,
\end{multline}
\begin{multline}
\label{KSasympt:subsol4}
\frac{\e{-4\tau/3}}{4\tau+1}
\begin{pmatrix}
3 \\ -1 \\ 12\tau \\ -4(4\tau+1) \\ \bzero_3
\end{pmatrix}
+ \frac{3\beta}{20(4\tau+1)} \e{-2\tau} 
\begin{pmatrix}
-(12\tau^2-51\tau-23) \\ -\frac1{24}(144\tau^2-56\tau+53) \\ 
2(56\tau^2+17\tau-4) \\ -5(4\tau+1)^2 \\ 
\bzero_3
\end{pmatrix} \\
+ \frac{3}{25(4\tau+1)} \e{-7\tau/3}
\begin{pmatrix}
\bzero_4 \\
-50 \\ -80\tau^2 -74\tau+49 \\ -80\tau^2+226\tau-51 
\end{pmatrix} \\
+ \frac{9\beta^2}{10240(4\tau+1)} \e{-8\tau/3} 
\begin{pmatrix}
 -1152\tau^3 +2720\tau^2 +7122\tau+1763 \\ 
 -384\tau^3 -992\tau^2/3 +650\tau +361/3 \\ 
 8256\tau^3 +10444\tau^2 -1392\tau -4169/4 \\ 
-3(320\tau^2+200\tau-343)(4\tau+1) \\ 
\bzero_3
\end{pmatrix}~,
\end{multline}
\begin{equation}
\label{KSasympt:subsol5}
\frac{\e{-2\tau}}{4\tau+1}
\begin{pmatrix}
4\tau+1/5 \\ 2\tau+23/30 \\ -4\tau-1/5 \\ 0 \\ \bzero_3
\end{pmatrix}
+ \frac{3\beta}{160(4\tau+1)} \e{-8\tau/3} 
\begin{pmatrix}
80\tau^2+144\tau+5 \\ \frac83(20\tau^2+36\tau+11) \\ 
-(80\tau^2+144\tau+5) \\ 0 \\ \bzero_3
\end{pmatrix}~,
\end{equation}
\begin{equation}
\label{KSasympt:subsol6}
\e{-7\tau/3}
\begin{pmatrix}
\bzero_4 \\
1/2 \\ -\frac{3}{50}(5\tau+4) \\ -\frac{3}{100}(10\tau-17) 
\end{pmatrix}
\end{equation}
and 
\begin{equation}
\label{KSasympt:subsol7}
\frac{\e{-8\tau/3}}{30(4\tau+1)}
\begin{pmatrix}
3(160\tau^2-172\tau+1) \\ -(160\tau^2+308\tau+121) \\ 
-6(260\tau^2-107\tau-16) \\ -450(4\tau+1) \\ \bzero_3
\end{pmatrix}~.
\end{equation}

\subsection{KS 7-scalar system: Small-$\tau$ behaviour}
\label{app:KSregsols}

In this appendix, we provide the independent small-$\tau$
behaviours. In the calculations, only the regular solutions are
needed. We include subleading terms up to and including order
$\tau^2$, as they are needed to avoid systematic errors in the initial
conditions. For completeness, we also list the singular solutions. 
In the following, we abbreviate 
\begin{equation}
\label{KSreg:A1def}
  A_1=A_1(0)=\frac12 3^{1/3}h(0)~.
\end{equation}
The seven regular small-$\tau$ solutions are
\begin{equation}
  \tau^2 
  \begin{pmatrix} 
  4 \\ 1 \\ 0 \\ -4 \\ -2 \\ -16A_1 \\ 0
  \end{pmatrix} 
  + \Order{\tau^4}~, \quad 
  \tau^2 
  \begin{pmatrix} 
  2 \\ 1 \\ (4/3)\tau \\ 0 \\ -2 \\ 0 \\ (4/3)\tau
  \end{pmatrix} 
  + \Order{\tau^4}~,  \quad 
  \tau^2 
  \begin{pmatrix} 
  13 \\ 3/2 \\ -(32A_1/3)\tau \\ 20 \\ 17 \\ 0 \\ (16A_1)\tau
  \end{pmatrix} 
  + \Order{\tau^4}~, \nonumber
\end{equation}
\begin{equation}
  \tau
  \begin{pmatrix} 
  0 \\ 0 \\ 1 \\ 0 \\ 0 \\ 0 \\ 1
  \end{pmatrix} 
  + \tau^2 
  \begin{pmatrix} 
  -3k^2/20+9/(80A_1) \\ -3k^2/40+9/(160A_1) \\ 0 \\ 0 \\ 
  3k^2/20 +11/(80A_1) \\ 0 \\ (1/6)\tau
  \end{pmatrix}
  + \Order{\tau^4}~, \quad
  \begin{pmatrix} 
  0 \\ 0 \\ 0 \\ 1 \\ 0 \\ 0 \\ 0
  \end{pmatrix} 
  + \tau^2 
  \begin{pmatrix} 
  -k^2/60+1/(120A_1) \\ -k^2/120 +1/(240A_1) \\ 0 \\ 
  k^2/6+1/(6A_1)  \\ k^2/60+19/(120A_1) \\ 0 \\ 0
  \end{pmatrix} + \Order{\tau^4}~,  \nonumber
\end{equation}
\begin{equation}
  \begin{pmatrix} 
  -3 \\ 1 \\ 0 \\ 0 \\ 0 \\ 0 \\ 0
  \end{pmatrix} 
  + \tau^2 
  \begin{pmatrix} 
  -2k^2/5 -3/(8A_1) \\ 13k^2/60 +1/(16A_1) \\ 0 \\ 0 \\
   -k^2/10+3/(8A_1) \\ 0 \\ 0
  \end{pmatrix} + \Order{\tau^4}~, \nonumber
\end{equation}  
\begin{equation}  
\label{KSreg:regsol7}
  \begin{pmatrix} 
  1 \\ 0 \\ 0 \\ 0 \\ 2 \\ 0 \\ 0
  \end{pmatrix} 
  + \tau^2 
  \begin{pmatrix} 
  k^2/5-1/10+31/(120A_1) \\ k^2/60-1/20+31/(240A_1) \\ 0 \\
  2/(3A_1)  \\ 3k^2/10-9/10+109/(120A_1) \\ 0 \\ 0
  \end{pmatrix} + \Order{\tau^4}~.
\end{equation}
The seven singular small-$\tau$ solutions are
\begin{equation}
  \begin{pmatrix} 
  0 \\ 0 \\ 1-8A_1/3 \\ 0 \\ 0 \\ 0 \\ 1
  \end{pmatrix} 
  + \Order{\tau}~, \quad 
\frac{1}{\tau}
  \begin{pmatrix} 
  1 \\ 0 \\ 0 \\ 2 \\ 2 \\ 0 \\ 0
  \end{pmatrix} 
  + \Order{\tau}~, \quad 
\frac{1}{\tau}
  \begin{pmatrix} 
  0 \\ 1 \\ 0 \\ 12 \\ 6 \\ 0 \\ 0
  \end{pmatrix} 
  + \Order{\tau}~, \quad 
\frac{1}{\tau}
  \begin{pmatrix} 
  0 \\ 0 \\ 0 \\ 12 \\ 1 \\ 4A_1 \\ 0
  \end{pmatrix} 
  + \Order{\tau}~, \nonumber
\end{equation}
\begin{equation}
\frac{1}{\tau}
  \begin{pmatrix} 
  0 \\ 0 \\ 4A_1\tau/9 \\ 1 \\ 0 \\ 0 \\ 0
  \end{pmatrix} 
  + \Order{\tau}~, \quad 
\frac{1}{\tau^2}
  \begin{pmatrix} 
  0 \\ 0 \\ 1 \\ 0 \\ 0 \\ -\tau \\ 1
  \end{pmatrix} 
+ \begin{pmatrix} 
  0 \\ 0 \\ -4A_1 k^2/3 +4A_1/9 +1/6 \\ 0 \\ 0 \\ 0 \\ 0
  \end{pmatrix} 
  + \Order{\tau}~, \nonumber
\end{equation}
\begin{equation}
\label{KSreg:singsol7}
\frac{1}{\tau^3}
  \begin{pmatrix} 
  1 \\ 1/2 \\ 0 \\ 0 \\ -1 \\ 0 \\ 0
  \end{pmatrix} 
+ \frac{1}{\tau}
  \begin{pmatrix} 
  0 \\ 0 \\ 0 \\ 0 \\ 0 \\ -4A_1k^2-22A_1/5-2/3 \\ 0
  \end{pmatrix} 
+ \begin{pmatrix} 
  0 \\ 0 \\ 16A_1k^2/3 +20A_1/3 +2/9 \\ 0 \\ 0 \\ 0 \\ 0
  \end{pmatrix} 
  + \Order{\tau}~.
\end{equation}
%


\section{KS spin-0 spectrum}
\label{KSspec}
Table~\ref{Table:KSextended} gives a somewhat more extensive collection of 
the spectrum we have found (we computed the spectrum up to $m^2=600$).

\begin{table}[h]
\begin{center}
\begin{tabular}{|r|r||r|r||r|r||r|r||r|r|}    \hline
$n$ & $m^2$ & $n$ & $m^2$ &  $n$ & $m^2$ &  $n$ & $m^2$ & $n$ & $m^2$ \\ \hline
 1 &    0.185 &  16 &    5.63 &  31 &   12.09 &  46 &   21.33 &  61 &   32.30 \\ [0mm]
  2 &    0.428 &  17 &    5.63 &  32 &   12.99  & 47 &   21.58 &  62 &   33.04 \\ [0mm]
  3 &    0.835 &  18 &    6.59 &  33 &   13.02  &48 &   22.10 &  63 &   34.82 \\ [0mm]
  4 &    1.28 &  19 &    6.66 &  34 &   13.31  &49 &   23.53 &  64 &   35.21 \\ [0mm]
  5 &    1.63 &  20 &    6.77 &  35 &   14.23  &50 &   23.95 &  65 &   35.54 \\ [0mm]
  6 &    1.94 &  21 &    7.14 &  36 &   15.03  &51 &   24.24 &  66 &   37.65 \\ [0mm]
  7 &    2.34 &  22 &    8.08 &  37 &   15.09  &52 &   25.94 &  67 &   38.17 \\ [0mm]
  8 &    2.61 &  23 &    8.25 &  38 &   16.16  &53 &   26.32 &  68 &   38.47 \\ [0mm]
  9 &    3.32 &  24 &    8.57 &  39 &   16.89  &54 &   26.67 &  69 &   39.32 \\ [0mm]
 10 &    3.54 &  25 &    9.54 &  40 &   17.03  &55 &   27.30 &  70 &   41.15 \\ [0mm]
 11 &    4.12 &  26 &    9.62 &  41 &   17.44  &56 &   28.95 &  71 &   41.79 \\ [0mm]
 12 &    4.18 &  27 &    9.72 &  42 &   18.61  &57 &   29.25 &  72 &   42.01 \\  [0mm]
 13 &    4.43 &  28 &   10.40 &  43 &   19.22  &58 &   29.62 &  73 &   44.22 \\  [0mm]
 14 &    4.43 &  29 &   11.32 &  44 &   19.40  &59 &   31.57 &  74 &   45.01 \\  [0mm]
 15 &    5.36 &  30 &   11.38 &  45 &   20.79 & 60 &   31.93 &  75 &   45.19 \\ [0mm]   \hline
\end{tabular}
\end{center}
\caption{Klebanov-Strassler spin-0 spectrum, first 75 values.} \label{Table:KSextended}
\end{table}


\section{Numerics}
\label{app:numerics}
In this section, we give a few details on how we computed the KS mass 
spectrum and how sensitive this spectrum is to numerical
uncertainties. (Not very sensitive at all, as we now argue.)
A quick summary is given in Table \ref{numericstable}.
The main program parameters are $\tau_{\rm IR}$, $\tau_{\rm UV}$, $\tau_{\rm
  mid}$ and $n_{\rm  imposed}$. Let us begin by explaining what they are.

\begin{table}[t]
\begin{center}
\begin{minipage}{7.7cm}
\begin{center}
\begin{tabular}{|c|c|c|c|c|} \hline
$\tau_{\rm IR}$ & $\tau_{\rm UV}$ & $\tau_{\rm mid}$ & $n_{\rm
  imposed}$& $m^2$   \\ \hline \hline
 0.001 & 20 & 1 & 500 & {\bf 0.18500} \\ \hline 
 0.2 & 20 & 1 & 500 & 0.18501 \\ \hline 
 0.001 & 11 & 1 & 500 & 0.18500 \\ \hline 
 0.001 & 25 & 5 & 500 & 0.18500 \\ \hline
 0.001 & 20 & 1 & 1 & 0.1850x \\ \hline   \hline
 0.001 & 20 & 1 & 500 & {\bf 1.28107} \\ \hline 
 0.2 & 20 & 1 & 500 & 1.28121 \\ \hline 
 0.001 & 11 & 1 & 500 & 1.28107 \\ \hline 
 0.001 & 25 & 5 & 500 & 1.28107 \\ \hline   
 0.001 & 20 & 1 & 1 & 1.285xx \\ \hline 
\end{tabular}
\end{center}
\end{minipage}
\begin{minipage}{7.7cm}
\begin{center}
\begin{tabular}{|c|c|c|c|c|} \hline
$\tau_{\rm IR}$ & $\tau_{\rm UV}$ & $\tau_{\rm mid}$ & $n_{\rm
  imposed}$& $m^2$   \\ \hline \hline
 0.001 & 20 & 1 & 500 & {\bf 4.12131} \\ \hline 
 0.2 & 20 & 1 & 500 & 4.12129 \\ \hline 
 0.001 & 11 & 1 & 500 & 4.12131 \\ \hline 
 0.001 & 25 & 5 & 500 & 4.12131 \\ \hline 
 0.001 & 20 & 1 & 1 & 4.12131 \\ \hline   \hline
 0.001 & 20 & 1 & 500 & {\bf 6.59516} \\ \hline 
 0.2 & 20 & 1 & 500 & 6.59519 \\ \hline 
 0.001 & 11 & 1 & 500 & 6.59516 \\ \hline 
 0.001 & 25 & 5 & 500 & 6.59516 \\ \hline 
 0.001 & 20 & 1 & 1 & 6.5951x \\ \hline 
\end{tabular}
\end{center}
\end{minipage}
\caption{A few sample mass values $m^2$ and their variation with some of the
  program parameters. $n_{\rm imposed}$ is the number of non-adaptive,
  imposed steps, within which we allow for adaptivity. 
``xx'' means the plot lacks detail to resolve further digits.
The conclusion is that even under strong changes of program
  parameters, roughly three significant digits remain stable.}
\label{numericstable}
\end{center}
\end{table}

The first two are self-explanatory: the actual radial interval of 
the Klebanov-Strassler background is
$\tau=[0,\infty]$, but to put the equations on a computer we need to
impose some cutoffs.\footnote{One could map the 
domain to e.g.\ $[0,1]$, but in simple examples we did not see any
improvement by doing so. Also, it might seem confusing
that we need to impose an IR cutoff when the background is smooth in
the IR. Just as in AdS, though, the {\it equations of motion} are singular for
$\tau=0$, but the singularity is regular. 
However, some of the regular boundary conditions coincide
for $\tau$ exactly zero, so to implement them faithfully
 we need $\tau_{\rm IR}>0$.} Thus, $\tau_{\rm IR}$ is an IR cutoff and 
$\tau_{\rm UV}$ is a UV cutoff. The parameter
$\tau_{\rm mid}$ is the midpoint for the midpoint determinant method
(see section \ref{method:method2}),
i.e.\ the position in $\tau$
where we compute
the matrix $\gamma_{ij}$. We will explain the  parameter $n_{\rm
  imposed}$ in a moment.

To integrate the 14 1st order ODEs between $\tau_{\rm IR}$ and
$\tau_{\rm UV}$, starting from analytically known boundary conditions given in
previous appendices, we 
use a standard adaptive Runge-Kutta 4th-5th order solver\footnote{see 
e.g.\ {\tt www.nr.com}}, implemented in {\tt Maple},
{\tt Fortran} and {\tt C}. For purposes of checking indiviual mass
values, the {\tt Maple} code is sufficient. For larger scans of
parameter space, the {\tt Fortran} and {\tt C} codes were useful. For
a typical run of $1,000$ values of $k^2$ with double precision, e.g.\ to generate 
the plots in fig.\ \ref{detplots}, the 
{\tt C} code takes about one hour on a 3 GHz Pentium 4 PC running Linux.

Many of the fields vary quickly in this system, and it turns out that the stepsize adjustment in
typical adaptive routines does not allow sufficiently fast change, so
we imposed a certain number of steps $n_{\rm imposed}$ that are not adaptive, and then
allowed adaptivity within those. Those steps were exponentially
spaced,
\begin{equation}
\tau_{i+1} = \left({\tau_{\rm UV} \over \tau_{\rm
    IR}}\right)^{\! {1 \over n_{\rm imposed}}} \tau_i~,  
\quad i=1, \ldots, n_{\rm imposed} \; .
\end{equation}
So $n_{\rm imposed}$ is a program parameter that when set to 1 reduces
the numerical integration to standard Runge-Kutta 4th-5th order over the entire range
$[\tau_{\rm IR}, \tau_{\rm UV}]$.

We performed some more extensive tests, but a few sample tests are shown in
Table \ref{numericstable}, along with the variation in the answer, the
mass value $m^2$. 

\begin{table}[t]
\begin{center}
\begin{tabular}{|c|c|c|c|c|} \hline
Method & Stable & Multifield & Avoids Exact & No fitting \\ \hline\hline
Standard shooting & $\times$ & $\times$ & UV & \checkmark \\ \hline
Blind shooting \cite[App.B]{Berg:2002hy} 
& $\times$ & $\times$ & \checkmark  & \checkmark \\ \hline
UV $\rightarrow$ IR Determinant & $\times$ & \checkmark & IR &
$\times$ \\ \hline
IR $\rightarrow$ UV Determinant & $\times$/\checkmark & \checkmark &
UV & $\times$ \\ \hline
Midpoint Determinant & \checkmark & \checkmark & $\times$ &
\checkmark \\ \hline
\end{tabular}
\end{center}
\caption{Sketch of numerical approaches. {\sl Stable} means numerical
  stability, in the empirical sense of Table~\ref{numericstable}, for the examples we
  studied. {\sl Multifield} means that it 
  generalizes easily to systems with many fields. {\sl Avoids Exact}
means one does not need exact analytical solution of approximations of
  the equations of motion in the UV or IR region, or both. {\sl No
  fitting} means one does not have to fit the numerical solutions to prescribed
  analytical behavior.}
\label{numcomparison}
\end{table}

So far, this all referred to the midpoint determinant method. We also tried some
other approaches, as outlined in Table~\ref{numcomparison}. Each of the other
approaches has some redeeming feature, but for computing the KS
spectrum, the midpoint method was superior. As can be seen from
the table, the main challenge in using the midpoint method is that one
needs to find approximate asymptotic analytical solution in both the
UV and the IR, but this can be done using symbolic
computation.\footnote{For example, {\tt Maple}
can output optimized {\tt Fortran} or {\tt C} code
implementing those solutions.} We would expect that for
backgrounds more complicated than KS, the midpoint method will remain 
the most efficient of those in Table~\ref{numcomparison}.

\end{appendix}

\bibliographystyle{JHEP}
\bibliography{bulk2}

\end{document}